\documentclass{emulateapj} 
 
\usepackage{natbib} 
\usepackage{graphicx} 
 
\slugcomment{to appear in ApJ}%
 
\begin{document} 
 
\shorttitle{Debris disks in NGC2547} 
\shortauthors{Gorlova et al.} 
 
 
\title{Debris Disks in NGC 2547}

 
\author{N. Gorlova$^{1,2}$, Z. Balog$^2$, G. H. Rieke$^2$, J. Muzerolle$^2$,
K. Y. L. Su$^2$, V. D. Ivanov$^3$, E. T. Young$^2$} 
\affil{$^1$Department of Astronomy, University of Florida, Gainesville, FL 32611-2055, {\it ngorlova@astro.ufl.edu}}
\affil{$^2$Steward Observatory, University of Arizona, 933 North Cherry Avenue, Tucson, AZ 85721-0065}
\affil{$^3$European Southern Observatory, Ave. Alonso de Cordova 3107, Casilla 19, Santiago 19001, Chile } 

 
\begin{abstract}

We have surveyed the 30 Myr-old cluster NGC 2547 for planetary debris disks using {\it Spitzer}.
At 4.5--8 $\mu$m we are sensitive to the photospheric level down to mid-M stars (0.2 M$_{\odot}$) 
and at 24 $\mu$m to early-G stars (1.2 M$_{\odot}$). We find only two to four stars with
excesses at 8 $\mu$m out of $\sim$400--500 cluster members, resulting in an excess fraction $\lesssim$1\% at this wavelength. 
By contrast, the excess fraction at 24 $\mu$m is $\sim$40\% (for B--F types).
Out of four late-type stars with excesses at 8 $\mu$m
two marginal ones are consistent with asteroid-like debris disks.
Among stars with strong 8 $\mu$m excesses one is possibly
from a transitional disk, while another one
can be a result of a catastrophic collision. Our survey demonstrates that the inner 0.1--1 AU parts
of disks around solar-type stars clear out very thoroughly by 30 Myrs of age.
Comparing with the much slower decay of excesses at 24 and 70 $\mu$m, disks clear from the inside out,
of order 10 Myr for the inner zones probed at 8 $\mu$m compared with a hundred or more 
Myr for those probed with the two longer wavelengths. 

\end{abstract}
 
 
 

 
 
\section{Introduction} 

The 5 to 50 Myr age range is critical for tracing the formation of terrestrial planets. 
At the beginning of this range, protoplanetary disks are ending their lives, the gas
has mostly escaped the system, and the remaining material is in the form of solid bodies
that do not yet approach planetary scale \citep{Meyer06,Hernandez05,Pascucci06,Sicilia07}.
By the end of the range, planetary embryoes thousands
of kilometers in size have formed in the inner few AU zone \citep{Chambers01, Kenyon06}. 
They are capable of stirring up smaller bodies and creating collisional cascades,
forming debris disks that might take a number of forms, such as a single ring \citep{Kenyon04a},
or of numerous rings in the presence of giant planets that gravitationally scatter
and trap dust into resonances \citep{MoroMartin02}. The Earth-Moon system was created
in a grand collision at this age \citep{Kleine02}. These debris systems are warmed by the central
stars and emit in the mid- and far-infrared. Therefore, by studying the infrared excess emission
of stars over this range of ages, we can characterize some features of the early evolution of terrestrial planets. 

NGC 2547 is an excellent target to examine the late stages of this critical period.
It is a rich open cluster in the Southern Hemisphere
at a distance of 400-450 pc \citep{Claria82,Naylor02}. \citet{Claria82} was first to carry out an extensive
photoelectric survey (in the $UBV$ bands) and determine its main properties.
He identified 70 members up to $V=13.4^{m}$ (B3 to $\sim$K0) and estimated the age to be
between 30 and 80 Myr from the brightest members at the Main Sequence (MS) turn off. Despite
lying only 8$\degr$ from the Galactic plane, the cluster was found to have only a small amount of reddening,
$E(B-V)$=0.06 $\pm$ 0.02.
\citet{Jeffries98} and \citet{Naylor02} extended the cluster sequence further 
down to $V=20^{m}$ (mid-Ms or $\sim$0.2 M$_{\odot}$, according to the models of
\citet{Siess00}),
by carrying out a $BVI$ survey within 15$\arcmin$ from the cluster center 
and identifying counterparts to \textit{ROSAT} sources.
Recently, \citet{Jeffries04} performed an $RIZ$ survey identifying members in the
substellar domain (down to 0.05 M$_{\odot}$).
\citet{Jeffries98} inferred an age of 10--20 Myr from isochronal fitting to the low mass pre-MS population,
at odds with the estimate of  $\sim$50 Myr from the cluster turn-off (though the latter is defined by only one or two stars).
The first measurements of lithium depletion in the K-M members again favored 50 Myr,
casting doubt on the validity of the pre-MS evolutionary models \citep{Jeffries03,Oliveira03}.
The age discrepancy was finally reconciled at 30 $\pm$ 5 Myr based on more precise observations
of X-rays with \textit{XMM-Newton} \citep{Jeffries06} and of the Li 
depletion boundary with the VLT \citep{Jeffries05}. The most recent $UBVRI$ study of  \citet{Lyra06}
confirms a 30--40 Myr age range after fitting four independent contraction tracks;
the distance and reddening from this study are on the low side of the previous range of
estimates (390 pc and $E(B-V)=0.03$). 

By comparing the disk population in NGC 2547 with clusters and field stars
of different ages we can verify whether indeed this age corresponds to a high level
of planetesimal activity, and whether the time-scale for inner planet formation differs among stars
of various masses. The \textit{Spitzer} survey of \citet{Young04} demonstrated the lack of dust near
the sublimation temperature for the bulk of members down to mid-K, while $\sim$200 K dust was
found for a quarter of the early-type members. An ambiguity remained for late K--M stars however,
where mid-IR excesses were marginally detected in a dozen stars. The uncertainty was partly due to
measurement errors, but also due to low number statistics of known members for the empirical
determination of the photospheric locus in the \textit{Spitzer} bands. 

The present study is a re-analysis of the \textit{Spitzer} data for NGC 2547 first presented in \citet{Young04}.
That study was performed early in the \textit{Spitzer} mission; we can now use improved image processing techniques
and flux calibrations, resulting in more precise photometry. Secondly, to identify debris disks we apply improved
color-selection criteria developed in the successive studies of other clusters, within the framework of the MIPS
open cluster survey \citep{Gorlova04, Gorlova06, Siegler06}. Thirdly, we use a new extended membership list,
allowing an improvement in the statistical significance of the results. We therefore can probe disks around the lower mass
stars and over the full area covered by \textit{Spitzer}. Finally, we report a multi-slit medium-resolution spectroscopic
survey of a subsample of photometric members. We improve the spectroscopic membership list for F-M stars
and confirm chromospheric emission in K and M members, a result that was somewhat ambiguous in previous studies. 

\section{Observations and Data Reduction}\label{red} 

\subsection{\textit{Spitzer} Photometry}

\textit{Spitzer} observed a $\sim 1\degr$-wide area centered on NGC 2547
with the IRAC and MIPS instruments at 3.6, 4.5, 5.8, 8.0, and 24 $\mu$m central wavelengths.
Observations of NGC 2547 with IRAC were obtained on Dec 19, 2003.
The high dynamic range mode survey with integration times of 0.4 s and 10.4 s covers about half 
a square degree centered on RA=08:10:13 and DEC=$-$49:13:25. The observation was repeated
twice in each position giving a total of 20.8 s integration time. The BCD frames from the SSC pipeline 13.2 were processed
and mosaiced using a custom IDL program provided by the IRAC team. We then performed photometry
for a new member list (\S\ref{memb}). Source finding and aperture photometry for IRAC were carried out
using {\tt PhotVis} version 1.10 which is an IDL-GUI based photometry visualization tool \citep{Gutermuth04}.
The radii of the source aperture, and the inner and outer sky annuli were at 2.4, 2.4 and 7.2 arc-second respectively.
We converted the MJy/sr units to DN/s using conversion factors 0.1088, 0.1388, 0.5952, and 0.2021 for channels
1, 2, 3, and 4 respectively. Then we calculated the standard IRAC magnitudes
using $-2.5\cdot log(flux_{DN/s})+zp$ where $zp$ is 21.9929, 21.2583, 19.0747 and 19.4372 for channels
1--4 respectively. The zero point term $zp$ includes: 1) the zero point of the flux
to magnitude transformation based on the large aperture
measurements of a standard star (19.66, 18.94, 16.88, 17.39 mag
for channels 1--4); 2) the aperture corrections
to account for the difference between the aperture sizes used for the standard star and for NGC~2547 photometry
($-0.21$, $-0.23$, $-0.35$ and $-0.50$ mag for channels 1--4);
3) 2.54 mag correction for exposure time in the mosaic. Originally we accepted all sources with photometric error less than 0.2 mag
as good detections, then we examined the detected sources visually to clean the sample of spurious
detections and non-stellar objects. However, because the IRAC images were obtained with only two dithers,
this procedure is limited at removing the effects of the cosmic rays. This issue is discussed further in \S \ref{iracexc}. 
We compared our photometry with the photometry of \citet{Young04} and found that the two datasets
agree within the errors. The IRAC counterparts of the known members of NGC 2547 were found
using a 2$\arcsec$ matching radius. 

The Multi-band Imaging Photometer for Spitzer (MIPS)
observed NGC 2547 at 24 $\mu$m on Jan 28, 2004 \citep{Young04}.
We used the image of \citet{Young04} but re-did the photometry. 
At 24 $\mu$m we performed PSF-fitting photometry; this approach is preferred
(compared to the aperture photometry we used with IRAC data) because of the good PSF sampling;
it also helps suppress the effects of the complex background at this wavelength. The procedure was identical
to that \citet{Gorlova06} initially applied to the Pleiades. Briefly, we first applied the {\sc iraf} package
{\tt daophot}, task {\tt phot} to obtain aperture photometry for the cluster members based on 2MASS coordinates.
We used these results as input into task {\tt allstar} with the template PSF constructed from 19 isolated
stars in our image with a range of brightness. The calibration of the PSF-fitting photometry is based
on a comparison of it with the aperture photometry of these isolated, well-measured stars.
To put instrumental magnitudes produced by {\tt allstar} into the ``Vega system'', we used the following
conversion factors: the aperture correction for an aperture radius of 3.5 px (1 pixel$=1.247\arcsec$)
and sky annulus from 5 to 10.5 pixels is 0.86 mag; 1 DN/s/px$=$ 1.63742E-06 Jy/px; $[24]=0^{m}$ is
equivalent to 7.14 Jy. The {\tt allstar} re-centered coordinates deviate by no more than 2$\arcsec$ from
2MASS coordinates, except for a few of the faintest members and bright extended objects. 

\begin{figure} 
\plotone{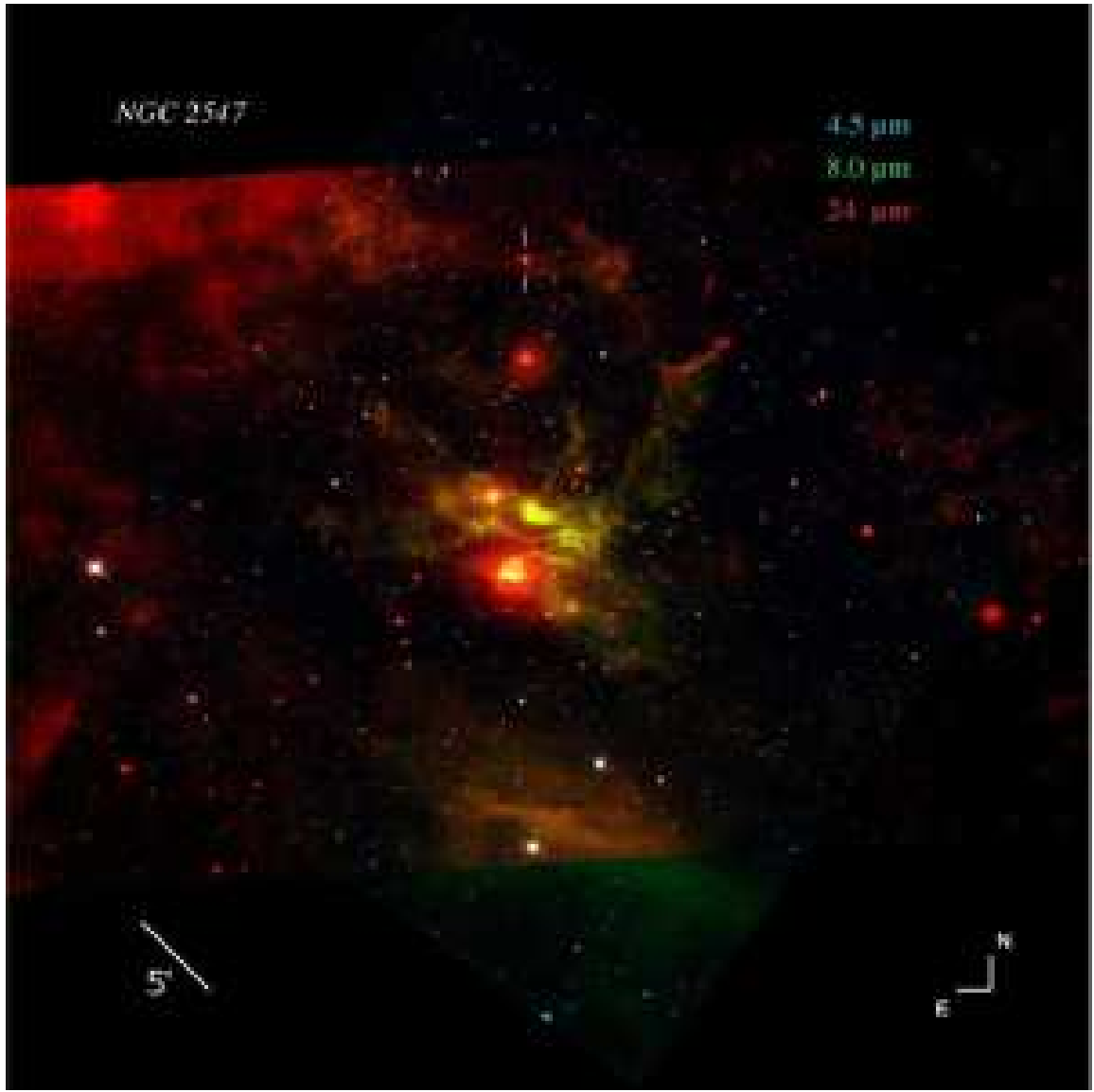}
\caption{NGC 2547 color mosaic composed of two IRAC bands and the MIPS 24 $\mu$m one.
The optically brightest members in the center of the cluster
are surrounded by extended 24 $\mu$m emission, a result of winds and heating of the interstellar cirrus.
They also excite the 7.7 $\mu$m aromatic feature, 
seen as the bright green area.
Two dashes indicate position of ID 9, a candidate low-mass member with a strong mid-IR excess.
}\label{fig1}   
\end{figure}

\begin{figure} 
\plotone{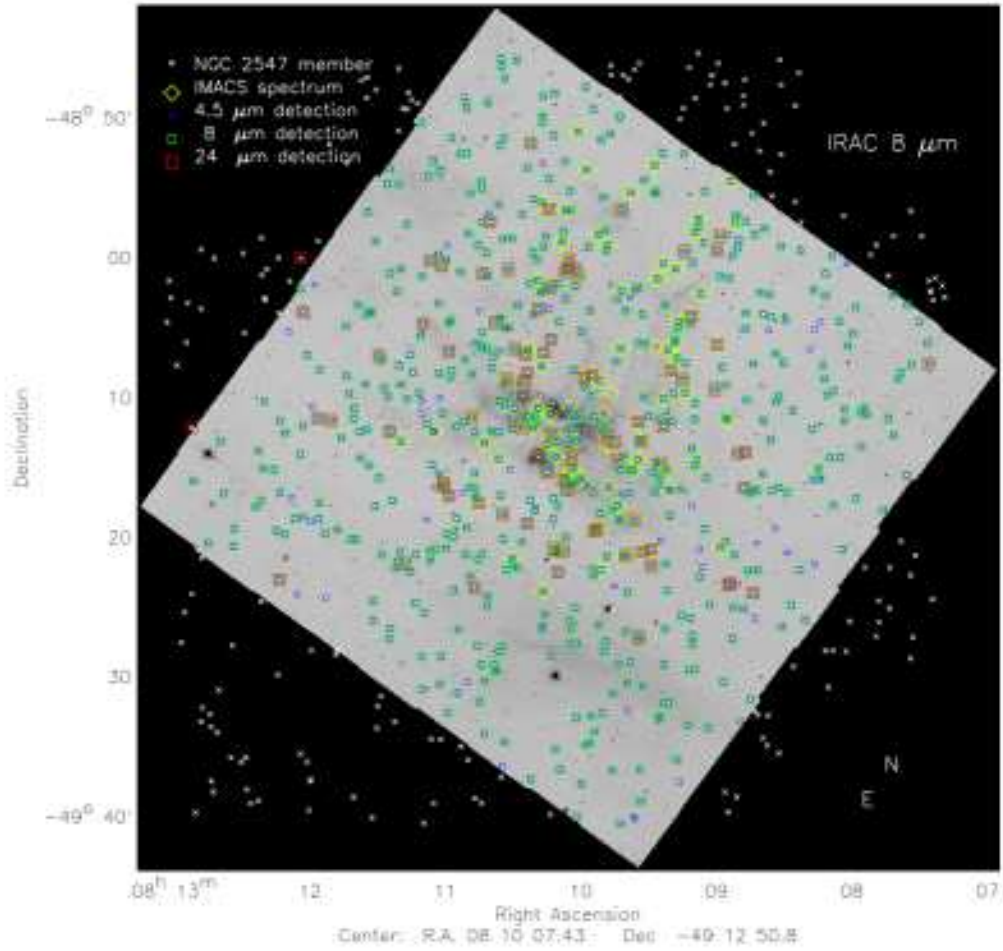}
\caption{IRAC 8 $\mu$m image showing area surveyed photometrically and spectroscopically
in this work. {\it Asterisks:} optically-selected photometric members from 
\citet{Naylor02} and \citet{Jeffries04}; {\it squares:} members detected with {\it Spitzer},
{\it diamonds:} sources for which we obtained spectra.
}\label{fig2}   
\end{figure}

Our \textit{Spitzer} photometry for optical candidate members
is given in Tables \ref{table1a} and \ref{table1b}.
Out of 860 objects we detect 606, 638, 589, 582, and
84 at maximum magnitudes 15.5, 15.7, 15.4, 15.0, and 11.7
in IRAC channels 1--4 and at MIPS 24 $\mu$m respectively.
We achieve 3--5 times more detections 
and go 1--2 mag deeper than \citet{Young04}.   
Figure \ref{fig1} shows the \textit{Spitzer} view of NGC 2547 composed of
the 4.5, 8 and 24 $\mu$m maps.
Fig. \ref{fig2} shows the 8 $\mu$m image onto which we
have overplotted the optical members,
those detected at 4.5, 8,
and 24 $\mu$m respectively, and the subsample we observed spectroscopically.
The IRAC images have a good match in depth and coverage 
to the optical surveys (except for a 0.2$\degr$ NE offset 
of the 3.5 and 5.8 $\mu$m channels from the cluster center).
The MIPS coverage is only slightly smaller in the NS direction,
but confusion with background sources and interstellar (ISM) nebulosity
limits its sensitivity to intermediate mass members
in the center of the cluster.

\subsection{Multi-slit Spectroscopy}

To evaluate the level of contamination
of our sample of photometrically selected members, as
well as to investigate the properties of excess sources, we have obtained
medium-resolution spectra in the red for a number
of candidate members, covering a wide range
of spectral types from B to mid-M.
These sources are marked by open circles in the
color-color plots in this paper.

The observations were performed in two runs in 2005 with the multi-object
spectrograph IMACS on the 6.5m Baade telescope (Magellan-1) at Las Campanas,
Chile. The atmospheric conditions were excellent with seeing typically around 0.6$\arcsec$.
We observed 3 fields with 5 masks. Two 15$\arcmin \times 15\arcmin$ fields were centered
on the cluster center and one was adjacent to the North-West (Table \ref{tablemasks}, 
Fig. \ref{fig2}). For the central fields we created separate masks for bright stars ($V=8-13^{m}$)
and faint stars ($V=13-20^{m}$) to avoid bleeding from the bright stars in long exposures. 
The faint masks contained 20--30 slits each. About a quarter of the spectra were rejected in
the final analysis due to overlap with other spectra, closeness to chip edges, or underexposure. 

We observed with the f/4 camera, 2$\times$2 binning (for higher S/N),
6$\arcsec$ $\times$ 0.7$\arcsec$ slits and the 1200 lines/mm grating, at the maximum
available angle of 32$\degr$.8 in 1st order. The spectra were dispersed across 4 chips
resulting in 3 gaps $\sim$100 \AA\,-wide. The aim was to target the region around the
Ca II IR triplet at 8500-8700 \AA\, at the maximum resolution. With a plate scale
of 0.22$\arcsec$/binned pixel, this resulted in resolution of 1 \AA\, (for 0.6$\arcsec$ seeing),
or $R=8,200$ (36 km/s) at the central wavelength of 8350 \AA\,. When reducing the data however we
discovered that the filter we used transmitted also in the 2nd order, contaminating the targeted red region
with 3600--4750 \AA\, flux. This affects mostly F-K stars that have strong metallic lines in the blue.
The effect is reduced at higher airmass. We alleviated the order overlap problem by carrying out
a spectral analysis relative to several templates chosen from the sample itself (as discussed in \S \ref{sp}),
and also with reference to external digital libraries that cover both spectral regions. 

We normally observed 3 science exposures, followed by quartz flat fields
and He-Ne-Ar arcs. Bias frames were obtained every other night.
The reduction was performed with the IMACS reduction package
{\sc Cosmos}, that performs alignment of the frames,
summation with cosmic ray rejection, flat-fielding,
wavelength calibration, sky subtraction
and 2-dimensional spectrum extraction. Final extraction was performed with the {\sc iraf} task
{\tt apall}.

In total, we extracted spectra for 88 sources, 84 of which are
in our member lists (Tables \ref{table1a}, \ref{table1b}).
2MASS 08093815-4918403 was resolved into 2 objects 1$\arcsec$ apart.
With seeing of 0.6$\arcsec$, the wings of the spectral PSFs overlap,
so we carried out their extraction with reduced aperture size.
The spectrum of the brighter source was given designation ``a'', and of the fainter one ``b''.
One more object was resolved into a pair -- 2MASS 08100398-4913027,
but only the spectrum of the brighter star was extracted.
The companion star is 1.3$\arcsec$ NE away, unresolved in 2MASS, and probably explains
the non-stellar flag in \citet{Jeffries04}. 

\subsection{Near-Infrared Spectroscopy}

For two solar-type members with strong \textit{Spitzer} excesses,
ID 8 \& 9, we obtained
K-band spectra targeting the CO band-head at 2.3 $\mu$m
to verify that these stars are not late-type interlopers.
The spectra were obtained with SofI (Son of ISAAC;
\citet{Moorwood98}) at the ESO 3.6m NTT telescope
on September 5, 2006. 
We used the medium resolution mode ($R=1300$)
with a 1$\arcsec$ wide slit.
The data were taken by nodding the target
along the slit. For each target
we obtained 4 images of 150 sec integration each. 

The data reduction consisted of the usual steps: flat fielding, sky
subtraction, and extraction of a one-dimensional spectrum from each
frame. Then, we wavelength calibrated the spectra with images of
Neon and Xenon lamps and combined the individual spectra into a
single spectrum. A B2.5 V star HIP 041781 and a G2 V HIP 43141 A 
were observed for telluric absorption correction.
Both targets and standards were observed at a narrow airmass
range of 1.4--1.5. Considering that the target stars
were expected to be of G--K type, to preserve
numerous weak atomic lines dominating their spectra,
we preferred to divide out telluric absorption with a B
rather than a G-type standard.
The K-band spectrum of a main sequence B star
is essentially featureless except for the
Br$\gamma$ line at 2.166 $\mu$m \citep{Hanson96},
which we removed by continuum interpolation
before division. A G-type telluric standard was reduced
in a similar way and served as a reference. 
The continuum shape of the final spectra
was straightened with a low-order polynomial,
to facilitate comparison of line strengths with spectroscopic standards
from the literature (Sec. \ref{trans}).

\section{Cluster Membership}\label{memb}

\subsection{Photometric Sample} 

\citet{Young04} used a sample of 184 stars from \citet{Naylor02} with membership
based on \citet{Dantona97} (DM97) isochrones, of which 169 were detected at least
in one IRAC band and 31 at 24 $\mu$m. \citet{Rieke05} report 24 $\mu$m measurements
for an additional 15 B-A members. Since then a deeper and wider survey has been carried
out in the $RIZ$ bands by \citet{Jeffries04}, that covers the central area within a 30$\arcmin$
radius from the cluster center, plus two fields 60$\arcmin$ away. Their $RIZ$ sample consists
of DM97 and \citet{Baraffe02} (BCAH02) - selected objects (their Tables ``A3'' and ``A2'' respectively).
In short, the selection process consisted of finding for each model a semi-empirical isochrone that best fits
color-magnitude diagrams of X-ray sources in NGC 2547 and the brown dwarfs in the Pleiades,
and identifying as members sources that lie 0.25 mag below and 0.9 mag above this isochrone,
which accounts for the photometric errors and unresolved binaries respectively.  
For our survey we decided not to give preference to either of the evolutionary
calculations and formed member sample by including all sources from both tables. The brightest stars
were saturated in this survey, we therefore complemented it 
with the ``enhanced catalog'' from \citet{Littlefair03} (``Table 1'') 
that is a combination of the \citet{Naylor02} $BVI$ ``wide catalog'' with the $UBV$ catalog of \citet{Claria82}.
The tables are available in electronic form at the Cluster Collaboration's Photometric Catalog
Page\footnote{{\tt http://www.astro.ex.ac.uk/people/timn/Catalogues/description.html}}. 

A number of $BV$-selected candidate members do not have $RIZ$ counterparts
in Table ``A2'' or ``A3'' of \citet{Jeffries04}. 
They are however listed in their Table ``A1'', 
which contains $RIZ$ photometry for everything in the NGC 2547 field.
After checking image quality flags for these sources,
we retained in the main member list (Table \ref{table1a})
those that were omitted apparently because of the saturation in the $RIZ$ bands. 
Those with non-stellar designation, flawed photometry due to bad pixels,
closeness to the CCD edge, or problematic background, as well as fainter sources
with good $RIZ$ photometry but not placed into tables ``A2'' and ``A3'' anyway,
were all moved into a separate table of less probable members (Table \ref{table1b}).
After removal of $\sim$4\% of sources without 2MASS counterpart within 2$\arcsec$,
the total number of objects in the two tables is 860.

\subsection{Spectroscopic Subsample}\label{sp}

\begin{figure} 
\epsscale{0.9}
\plotone{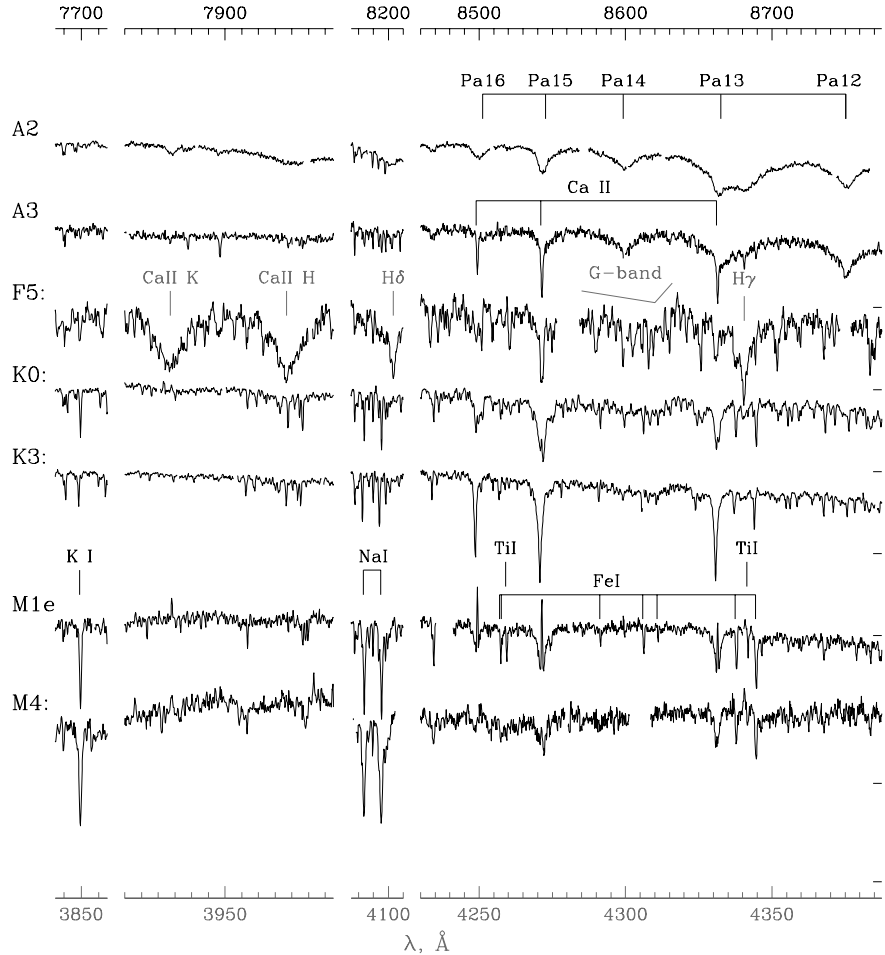}
\caption{Spectroscopic sequence in NGC 2547 based on the representative IMACS spectra
of non-excess stars.
All spectral segments have been normalized to unity
at the short-wavelength end and shifted vertically for clarity;
horizontal dashes on the right mark the level of the zero flux for each spectrum.
Spectral features marked in \textit{black} with the wavelength scale
given at top correspond to the red light (flux transmitted in the 1-st grating order).
Features marked in \textit{grey} with the wavelength scale on bottom
correspond to the blue light (unblocked flux from the 2-nd order). 
Distance between ticks: 10 \AA\, on the top scale and 5 \AA\, on the bottom one.
The wavelength scale is heliocentric. 
Line identifications are from \citet{Munari99}, \citet{Torres93}, and the Digital Spectral
Classification Atlas of R. O. Gray.
Examples of fast and slow rotation are given for A types,
and with and without emission in the Ca II lines for K and M types. 
Colon means that spectral type was estimated based on the $V-K$ color.
From top to bottom: 2MASS08092958-4906166, 2MASS08095066-4912493,
2MASS08102774-4912095, 2MASS08101292-4914094,
2MASS08101944-4907444,
2MASS08093685-4914214, 2MASS08092592-4909585.
}\label{fig3}   
\end{figure}

\begin{figure} 
\epsscale{0.9}
\epsscale{0.9}
\plotone{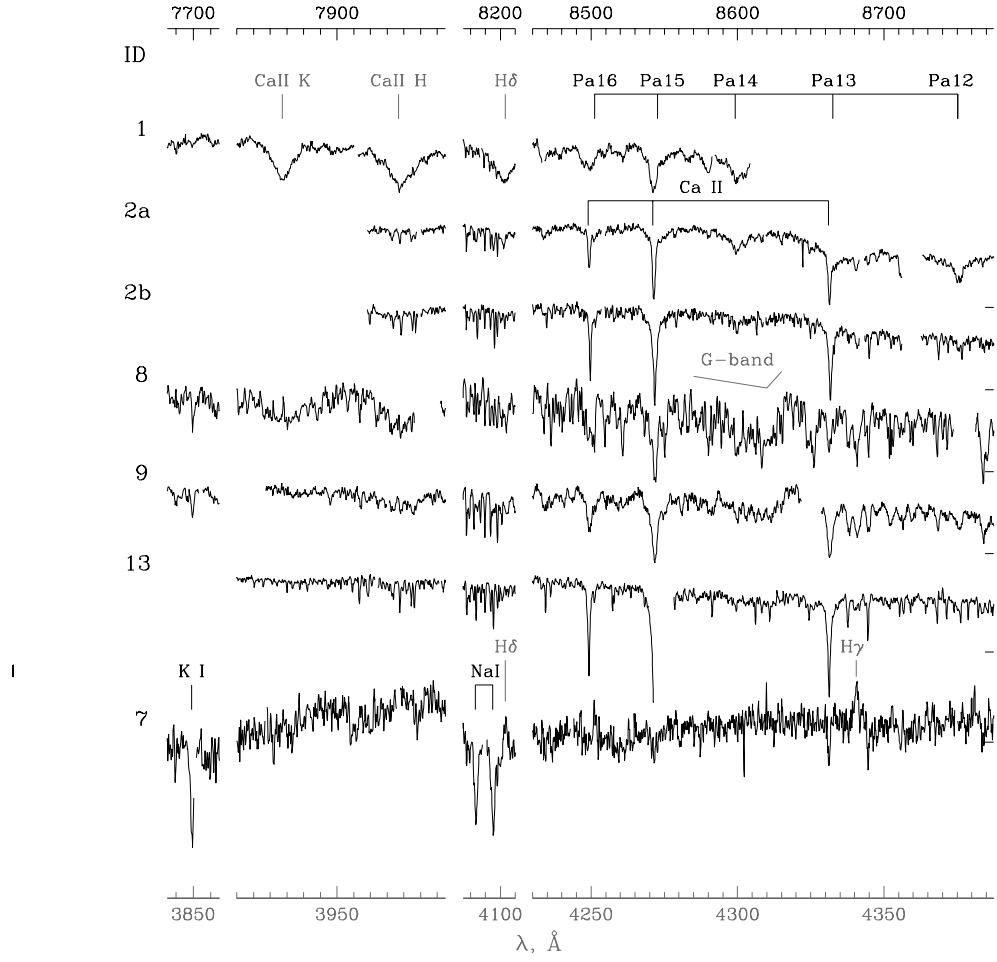}
\caption{IMACS spectra of some excess stars from Table \ref{tableexc}. 
From top to bottom: 2MASS08091401-4904029,
2MASS08093815-4918403 (resolved into two components), 2MASS08090250-4858172, 2MASS08101691-4856291,
2MASS08100961-4915540, 2MASS08093547-4913033. They compare well with spectra of non-excess
members in Fig. \ref{fig3}. 
}\label{fig4}   
\end{figure}

Because of the unfortunate overlap of the blue and red regions
in our spectra, spectral classification is not trivial.
To interpret the spectra we first plotted them
along a sequence in $V-K$ color, assuming that members would form
a monotonic sequence of spectral line strengths from which excess members and 
interlopers might stand out.
Fig. \ref{fig3} shows representative spectra along this sequence,
while Fig. \ref{fig4} shows spectra of excess candidates (\S\ref{exc}).
The segments shown contain prominent features used in spectral classification.

We found variability in the strength of the blue spectrum
that correlates with the airmass of observations.
The February observations were obtained at a slightly larger airmass
and the blue lines are weaker than in December.
As a result, we can not accurately differentiate between late F to K classes
based on the strength of such
classical indicators as Ca II H \& K, Balmer lines and the G-band.
The IR Ca II triplet itself does not vary significantly between F--K types \citep{Allen95,Carquillat97},
while the weaker Fe I and Ti I lines get mingled with metallic lines from the blue.
On the other hand, we can confidently identify B--A types based on the presence of the Pashen series,
and M types based on the K I and Na I strong lines in the red. Our analysis
significantly expands the spectral characterization of the cluster population, which previously
concentrated on the brightest B--A members
and a few K--M stars for the Li studies.

To measure radial velocities, we split our spectroscopic sample
into four groups that we designate ``B--F'', ``G'', ``K'' and ``M'',
with 13, 13, 40 and 23 objects respectively.
The names of these radial velocity groups
only roughly represent spectral types, due to variations
in the visibility of the blue spectrum. 
Within each group we picked a template to use for cross correlation,
requiring only for it to be of high S/N and contain no
chip gaps or bad pixels in the main features of interest.
These templates are 2MASS 08095066-4912493, 08095109-4859022,
08101944-4907444, and 08092437-4906282 respectively.

We measured the radial velocities of these templates using 3--6 
lines of Ca II, Fe, Ti in the 8500--8700 \AA\, interval
with line identifications from \citet{Munari99}.
We obtained 27.4, 58.0, -20.6 and 21.1 km/s with an RMS uncertainty
of few km/s.
We then used the {\sc iraf} task {\tt fxcor} to correlate
the template spectra with the objects in each group in the same 8500--8700 \AA\, region,
and converted the relative wavelength shifts into heliocentric velocities.
\citet{Robichon99} reports the radial velocity for the cluster to be 14.4 $\pm$ 1.2 km/s based on 5 stars,
\citet{Claria82} 15.4 km/s based on 10 stars,
and \citet{Jeffries00} 12.8 km/s $\pm$ 0.9 based on 20 stars.
Since we did not observe radial velocity standards, to put our measurements on the absolute scale,
we averaged our values in the M group (for which we achieved the highest accuracy)
to obtain 21.7 km/s (with RMS$=\pm$3.5 for 22 stars) and then subtracted
8.9 km/s from all our values to put them on the same scale
as in \citet{Jeffries00}.

\begin{figure} 
\plotone{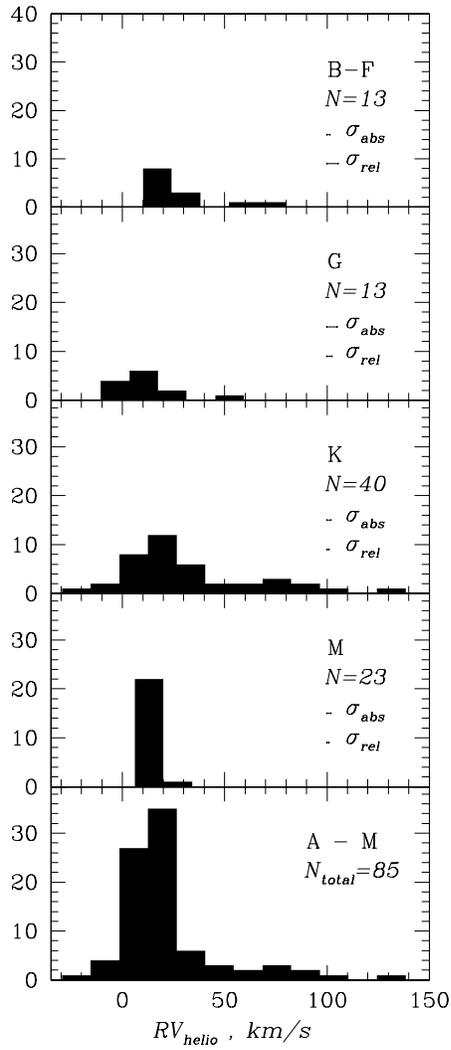}
\caption{Distribution of radial velocities in NGC 2547 measured from IMACS spectra
relative to the respective template in each spectral group. Contamination of K type
members by background giants can be seen from the secondary peak at 75 km/s.
$N$ gives the number of spectra in each group.
}\label{fig5}   
\end{figure}

The histograms of these adjusted radial velocities
for each radial velocity group and for all the groups together are plotted in
Fig. \ref{fig5}, while the individual values are listed
in Table \ref{tablesp}.
Two 1 $\sigma$ error bars are shown in each panel of Fig. \ref{fig5},
$\sigma_{abs}=$2--5 km/s being the uncertainty of the standard star velocity determination,
and $\sigma_{rel}=$1--6 km/s an average uncertainty of the relative velocity
between the standard and a given star in the group.   
The first error gives the limit on the
shift of the entire distribution along the x-axis,
while the second error contributes to the width of the distribution.
$\sigma_{abs}$ is taken to be the RMS deviation from the mean
when measuring velocity of the standard from several lines;
it compares well with the uncertainty of the wavelength calibration
of the arc lamp spectrum.
$\sigma_{rel}$ is calculated by {\tt fxcor} based on the goodness of the match
of the target and the shifted standard spectra;
it only exceeds $\sigma_{abs}$ in the BF group, due to
the cross-correlation of the broader hydrogen lines.
The larger $\sigma_{abs}$ of the G standard
is due to strong blending with the blue spectrum,
which makes it difficult to identify clean lines for the radial velocity determination.
A secondary peak at $\sim$75 km\,s$^{-1}$ in the K group
is real and most certainly arises from contaminating background stars, as one might expect in this spectral range.
\citet{Naylor02} estimated that the contamination
of the $BVI$ sample by background stars is negligible,
except in the 14.0 $<V<$ 15.5 range where the cluster
sequence crosses the ``finger'' of Galactic red giants
and contamination is predicted to be as high as 40\%.
The combined histogram at the bottom panel does not include the 4 earliest type
stars with the biggest $v\,sin\,i$ and discrepantly large radial velocities.
It can be used to identify as potential non-members stars with radial velocity $\gtrsim$50 km/s.
None of them appears among the 8 $\mu$m excess objects, and only one has
a possible 24 $\mu$m excess (2MASS 08101474-4912320).

\subsection{Proper Motions}\label{pm}

As an additional constraint on membership, we analyzed proper motions
for our photometric sample. The UCAC2 catalog and its bright stars supplement
\citep{ucac2}\footnote{Also available
via the VizieR Online Data Catalogs I/289 and I/294.}
contain measurements for all NGC 2547 B--F stars and 86\% of G--K stars,
but only for 6\% of M stars (when using 
a coordinate matching radius of 2$\arcsec$)\footnote{Spectral classes
are assigned based on $(V-K)_{0}$
or on $(R-K)_{0}$ color when the former is not available,
following \citet{Bessell88} and \citet{Kenyon95} dwarf calibrations.
Reddening $E(B-V)=0.06$ and following relations were adopted for all NGC 2547 stars:
$A_{V}=3.1E(B-V)=A_{R}/0.748=A_{K}/0.112=A_{[5.8]}/0.048=A_{[8]}/0.048=A_{[24]}/0.0$}.
We fitted each distribution of proper motions (in RA$*$cos(Dec) and Dec) with a Gaussian
to get a mean of $-$4.6 mas/yr and 5.7 mas/yr,
and $\sigma$ of 5.0 and 4.8 mas/yr, respectively.
Objects with proper motion vectors within 10 mas/yr (2$\sigma$)
of the cluster mean have been marked as members (flag "Y"
in the last column of Tables \ref{table1a} and \ref{table1b}). Objects more than
10 mas/yr from the mean are marked as proper motion non-members.
We find the proper-motion-confirmed member fraction to be 95\%, 81\%, 52\%, 77\% and 80\%
among BA, F, G, K, and M-type photometric members, respectively. 

\section{Identifying Excess Candidates}\label{exc}

\subsection{IRAC Excesses}\label{iracexc}

\begin{figure} 
\plotone{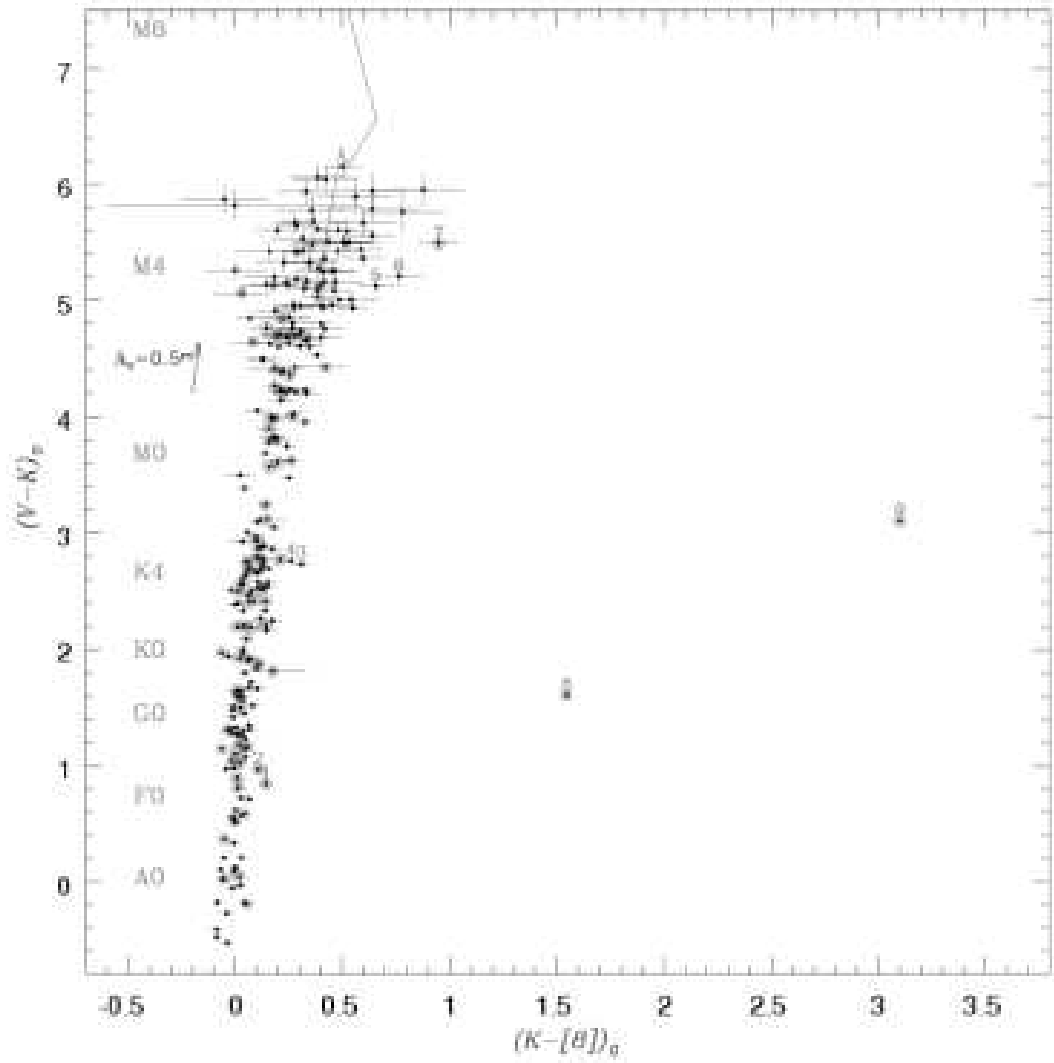}
\caption{Color-color diagram (de-reddened) for identifying 8 $\mu$m excesses
using $V-K$ color to trace photospheric flux. {\it Solid circles:}
$BVRIZ$-selected NGC 2547 members (from Table \ref{table1a}); {\it triangles:} $BV$-selected
possible members (from Table \ref{table1b}); {\it open circles:} spectroscopically observed (from
Table \ref{tablesp}). The solid line connects seven M3--M6.5 field dwarfs from \citet{Patten06},
$V$ mags are taken from VizieR. 
}\label{fig6}   
\end{figure}

The four IRAC bands spanning 3-8 $\mu$m
trace dust from the near-sublimation temperature of $\sim$1000 K to about 400 K,
corresponding to radii within 1 AU for solar-type stars.
To identify possible
8 $\mu$m excesses from the IRAC data, we plot sources
from Tables \ref{table1a} and \ref{table1b} on the $V-K$ vs $K-[8]$ diagram 
in Figure \ref{fig6}. The large baseline of the $V-K$ color results in a monotonic
relationship between increasing redness and increasingly late stellar spectral type. 
For stars with no excess, the similarity of the photospheric colors for a given
spectral type results in a narrow locus, extending from $V-K \sim 0$ (for A-type and
earlier stars) to 5 (early M); the scatter in this locus increases for $V - K > 5$
due to measurement errors. Stars to the right of this locus are candidates to have
excesses at 8 $\mu$m.

We visually examined the quality of the image of each object with an apparent 8 $\mu$m excess.
We removed nine objects from the plot because of a nearby companion, cosmic ray hit,
artifact from a bright star, or a non-uniform nebulosity contaminating the 4 px-wide aperture.
One more object (2MASS 08104684-4927452) was rejected because
it was too faint on the $V$ vs $V-K$ diagram.
Four more were not plotted because they had non-stellar flags on the $RIZ$ magnitudes.
Only one of these 14 objects is detected at 24 $\mu$m (2MASS 08100961-4915540), and its
24 $\mu$m excess is in question
as well, due to strong nebulosity.
All these sources were omitted from Fig. \ref{fig6}.

\begin{figure} 
\plotone{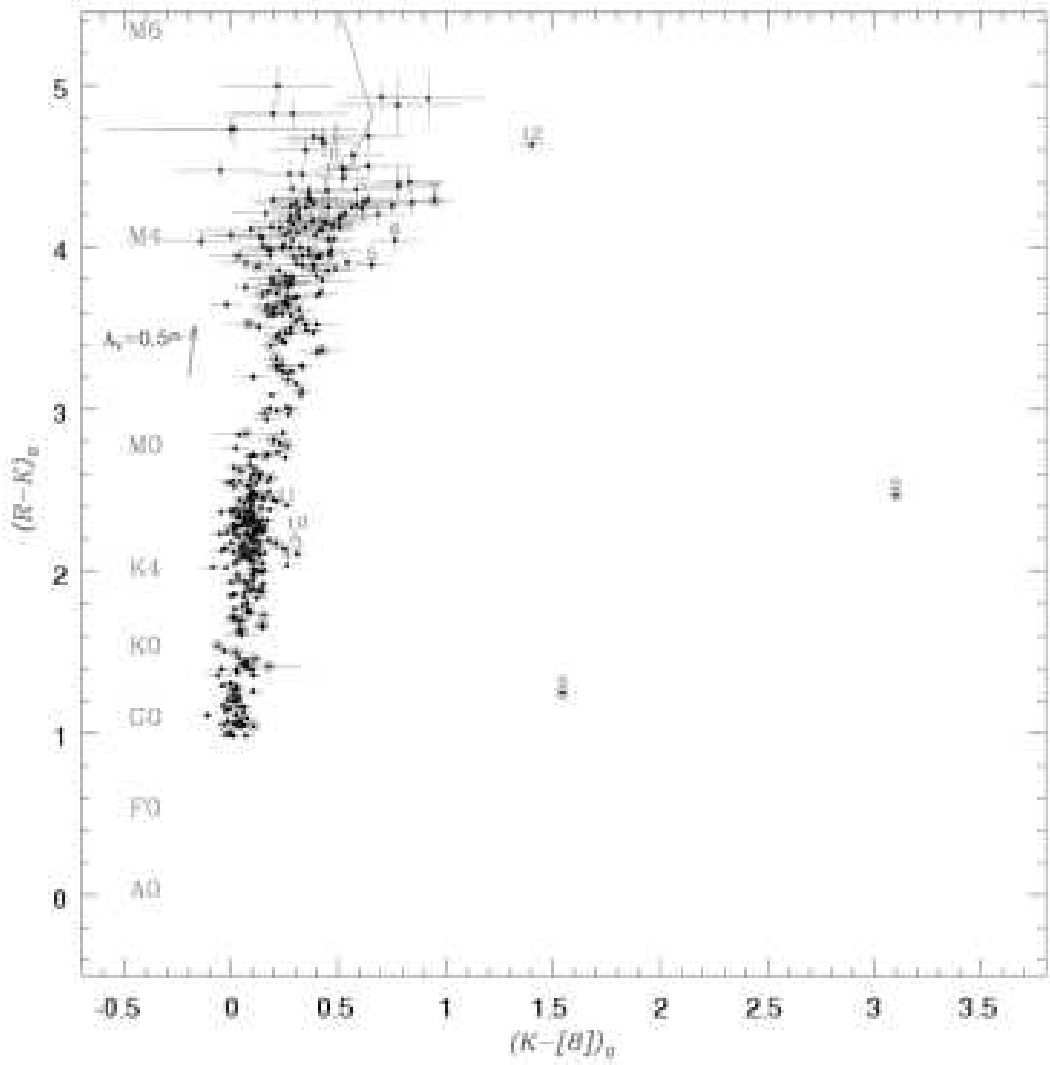}
\caption{Color-color diagram for identifying 8 $\mu$m excesses
using $R-K$ color to trace photospheric flux. Symbols same as in Fig. \ref{fig6}.
$R$ mags for field dwarfs of \citet{Patten06} 
are derived from $V-R$ vs. SpT relation of \citet{Kenyon95}.
}\label{fig7}   
\end{figure}

A similar diagram was constructed in the $R-K$ color (Fig. \ref{fig7}). 
We omitted the objects excluded from Fig. \ref{fig6}
plus nine additional objects (none of which has a $V$ mag) with questionable 8 $\mu$m excesses
due to poor image quality.
One more (2MASS 08111674-4906564) was rejected because
it is too faint on the $R$ vs $R-K$ diagram.
Of these 10 objects only one is detected at 24 $\mu$m, with only marginal excess
due to closeness to the detection limit (2MASS 08090351-4909148).
Tables \ref{table1a} and \ref{table1b} include footnotes regarding source selection issues.

\begin{figure} 
\plotone{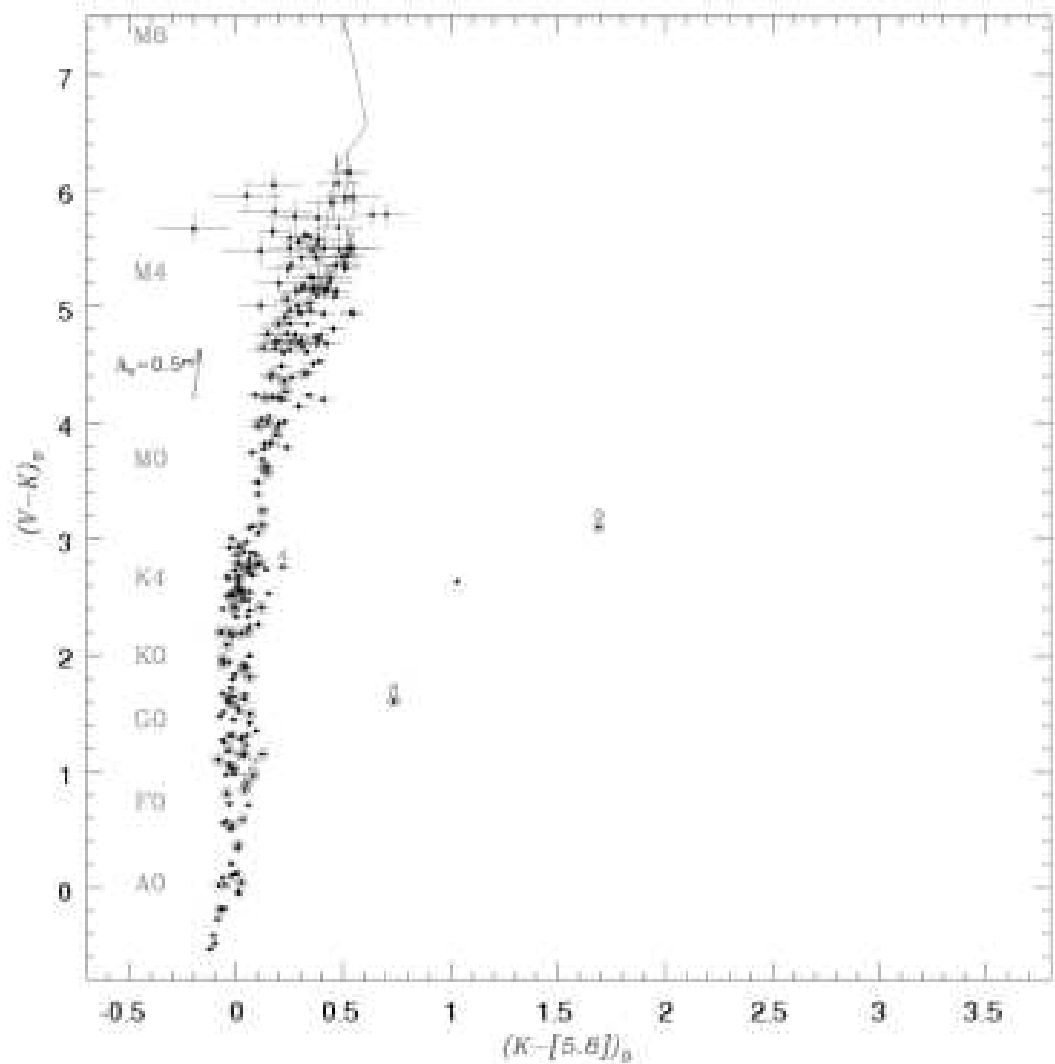}
\caption{Color-color diagram to confirm 8 $\mu$m excesses in the neigboring
5.8 $\mu$m IRAC band. Symbols same as in Fig. \ref{fig6}.
}\label{fig8}   
\end{figure}

\begin{figure} 
\plotone{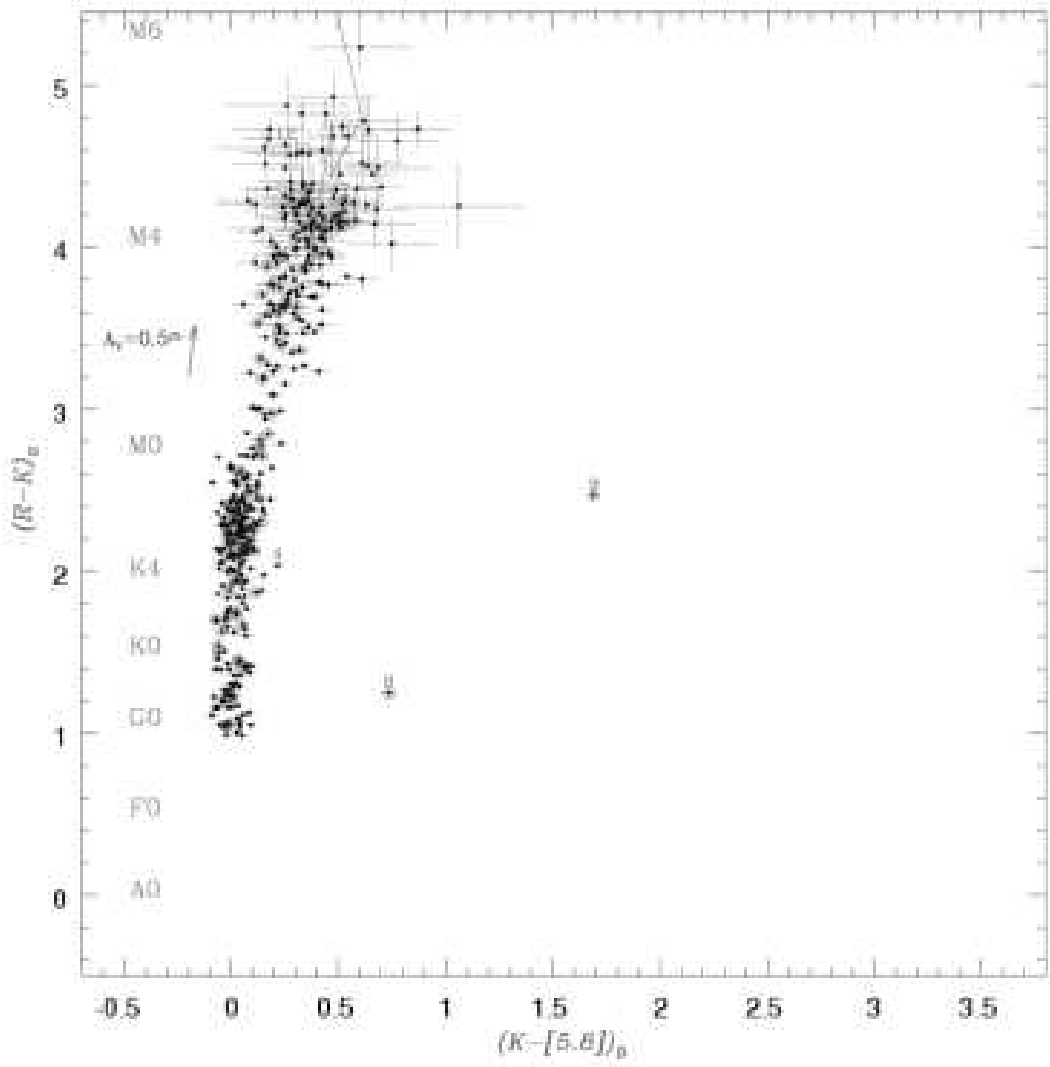}
\caption{Similar diagram as in Fig. \ref{fig8} but with $R-K$ color.
}\label{fig9}   
\end{figure}

The remaining 8 $\mu$m excesses were marked with numbers on Figures \ref{fig6} and \ref{fig7}.
Excess objects were visually selected by being $\geq$3 $\sigma$ away from the
average locus formed by the majority of members of the corresponding $V-K$ ($R-K$) color.
The 8 $\mu$m channel is significantly affected by the cosmic ray hits. Since the size of the IRAC beam
is only about 4 pixels, for faint objects
it is sometimes difficult to decide whether a detection is real or an artifact of a cosmic ray hit.
However, if an excess is found at 8 $\mu$m,
we can expect to detect it also in the 5.8 $\mu$m band.
We therefore constructed $V-K$ vs $K-[5.8]$, $R-K$ vs $K-[5.8]$ diagrams
(Figs \ref{fig8}, \ref{fig9}). As can be seen from these plots,
most faint members with 8 $\mu$m excesses do not stand out
at 5.8 $\mu$m: ID 3 (2MASS 08104343-4930158),  5 (2MASS 08085856-4911171), 6 (2MASS 08101041-4858052), 
7 (2MASS 08093547-4913033), 10 (2MASS 08104401-4859372),
11 (2MASS 08083389-4907176), 12 (2MASS 08094287-4903413).
To guard against bad photometry due to cosmic ray hits, we required
an excess in faint sources to be apparent at both 5.8 and 8 $\mu$m. 
In the absence of additional information about objects rejected
based on this requirement we do not discuss them further,
with the exception of ID 7, which was already noticed by \citet{Young04} 
to have a 24 $\mu$m excess and for which we obtained a spectrum.
Properties of ID 7 together with the remaining 8 $\mu$m excess
objects are presented in Table \ref{tableexc}.
 
\subsection{MIPS Excesses}

The cluster has mass segregation, with members
above 2 $M_{\odot}$ (mid-Fs) all gathered within the central 20$\arcmin$
\citep{Littlefair03, Jeffries04}. This mass limit coincides with the limit for photospheric
detection at 24 $\mu$m,
which is at $\sim$10.8$^{m}$, or 0.3 mJy (where the distribution
of [24] magnitudes experiences a sharp downturn).
We can achieve excess \textit{detections} to lower masses,
but we will be unable to calculate the 
excess \textit{fraction} (the fraction of stars with excess) there.
Thus, compared to \citet{Young04}, we are probing the disk fraction at 24 $\mu$m to a similar
(stellar) mass limit. The advantage of our study
is improved statistics for excess detections down to early M stars,
due to the larger area covered; as a result,
we detect a few sources with excesses even more extreme than in \citet{Young04} (\S \ref{prop}).

\begin{figure} 
\plotone{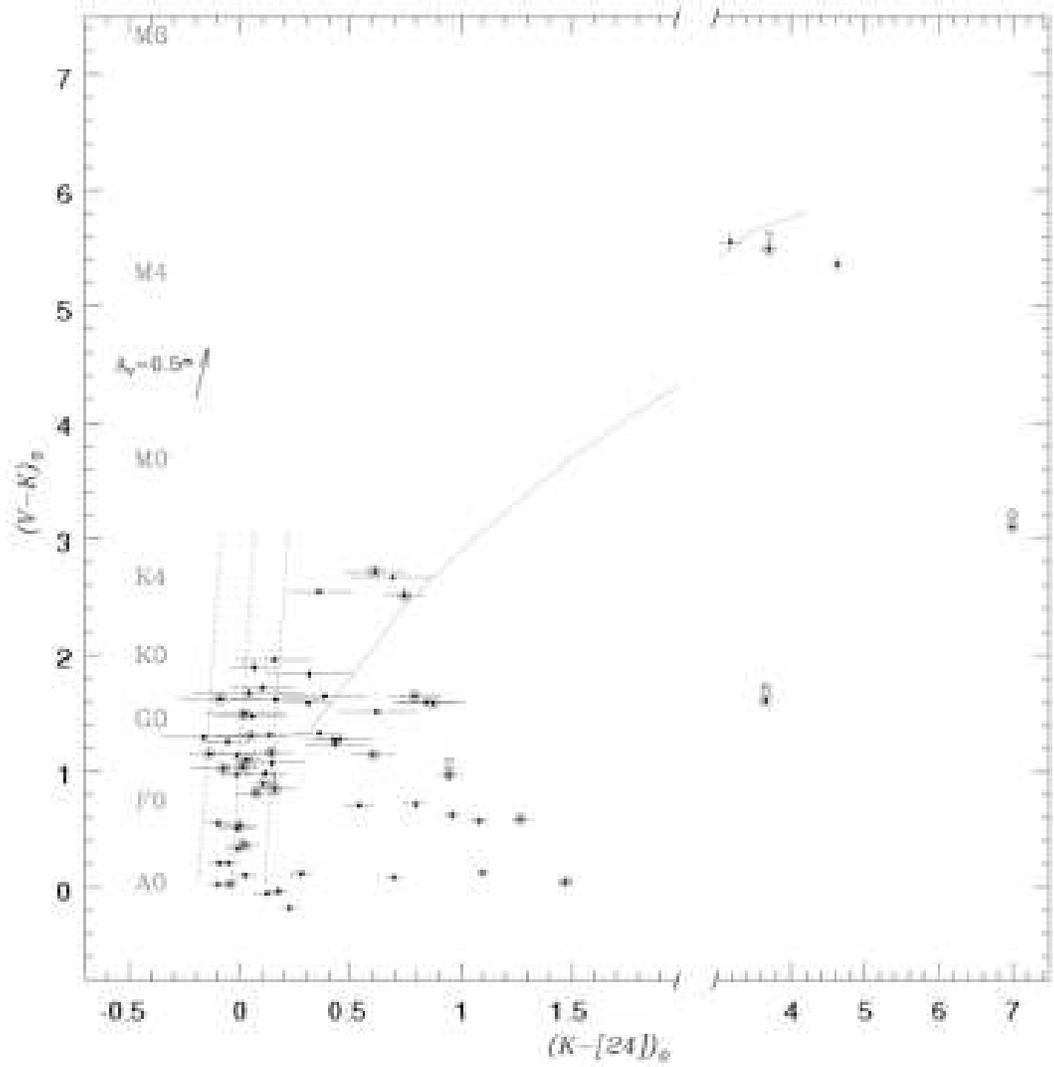}
\caption{Color--color diagram used to identify 24 $\mu$m excesses.
The solid gray line represents our detection limit at 24 $\mu$m.
It approximates the locus of the non-detected members by assuming their $[24]<10.8^{m}$,
and so is a reflection of the $K$ vs $V-K$ sequence. It shows that we are complete
at 24 $\mu$m up to $V-K \sim$1.3 (late Fs).
Objects to the left of this line are either binaries or
situated in the very clean from nebulosity region.
Dashed lines are the non-excess locus $\pm$ 3 $\sigma$ derived for the Pleiades stars
(from \citet{Gorlova06}), which we utilize here to identify excesses in NGC 2547. 
Symbols are the same as on the previous figures. The triangle at (4.62, 5.37) is 2MASS08090344-4859210
discussed in \S \ref{prop8}.
}\label{fig10}   
\end{figure}

\begin{figure} 
\plotone{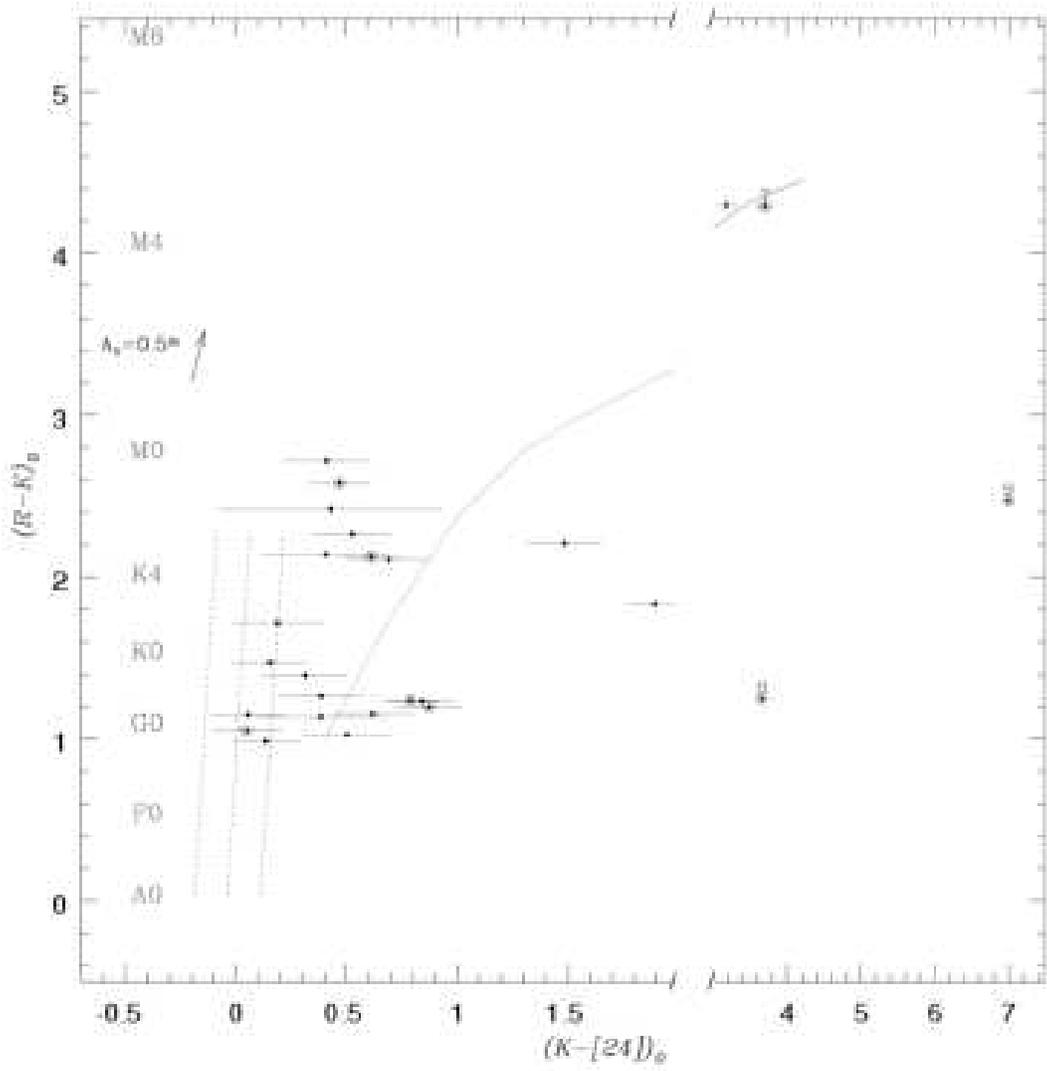}
\caption{Same diagram as in Fig. \ref{fig10} but with $R-K$ color.
}\label{fig11}  
\end{figure}

How do we identify 24 $\mu$m excesses?
The cluster is young and has a continuum of objects
with $K-[24]$ bigger than $\sim$0 expected from the Rayleigh-Jeans tail
of the pure photospheric emission for stars earlier than early M (Figs. \ref{fig10}, \ref{fig11}).
The non-excess locus thus cannot be easily identified from the stars of NGC 2547 itself.
We therefore use the Pleiades locus as a fiducial.  
One can see that all the NGC 2547 stars fall inside or to the right of the
adopted non-excess boundaries relative to this locus.
We regard objects 3 $\sigma$ to the red relative to the
central line to have a 24 $\mu$m excess and report them in Table \ref{tableexc}.
To reproduce the Pleiades locus in the $R-K$ color in Fig. \ref{fig11},
we used the relation between $R-K$ and $V-K$ for B9--K7 MS stars
from \citet{Kenyon95}. Two new excesses are identified on this diagram, with
$K-[24]$ between 1.5 and 2 mag.

The five optically brightest members (B stars) formally show excess at 24 $\mu$m,
but in all of them the emission is extended beyond the 6$\arcsec$-wide MIPS beam,
corresponding to 2400 AU at 400 pc.  This scale is too big for a typical
debris disk. The behavior was first encountered in follow-up studies of 
$IRAS$-discovered excesses, and was dubbed 
``the Pleiades phenomenon''.  It is explained by accumulation of matter
around massive stars as a result of either mass loss or
the radiation pressure exerted on the interstellar dust,
and should not be confused with a debris disk that results from
the shattering of planetesimals.
Indeed, two bubbles are seen on Fig. \ref{fig1} around these ``halo'' stars.
The bigger one in the northern part of the cluster is centered
on the B5 II member HD 68451, and the more compact one in the center of the cluster
is clearly produced by the B3 IV star HD 68478.
We mark these stars in Table \ref{table1a}
and omit them from the $K-[24]$ plots.  

At magnitudes near the detection limit one has to worry about confusion
with ISM clumps and extragalactic objects.
We examined the images of all 24 $\mu$m excess sources.
To help the discussion and guide the eye on Fig. \ref{fig10}, we split these objects
into several groups based on $V-K$ (see Table \ref{tableexc}).
Three of the four excess sources with $V-K\sim1.6$ and $K-[24]=0.6-0.9$
have distorted shapes, while 6 sources out of 8 with $V-K>2.5$ ($R-K>1.8$),
$K-[24]=0.6-4$ are at the level of fluctuations in the nebulosity.
We therefore call these excesses ``possible''.
We note that the region of $V-K=2-4$ is also where contamination from
background giants is strongest \citep{Littlefair03}. Indeed, 2MASS 08101474-4912320,
whose membership is only based on $BVI$ photometry, has a radial velocity $>100$ km/s
which is much larger than the cluster mean (\S \ref{sp}).
We do not discuss objects from this group any further.
The images of the remaining excess sources however (including the four most
extreme ones with $K-[24]>3.5$) appear symmetrical,
and where radial velocities are measured,
they are consistent with cluster membership. We discuss them in \S \ref{prop}.

\section{Properties of the Excess Candidates}\label{prop}

\subsection{8 $\mu$m excesses}\label{prop8}

Sources with excesses at 8 $\mu$m are rare in this cluster. 
We discuss the characteristics of each of them below.

\textit{ID 1} has only a marginal excess at 8 $\mu$m and has a border-line one at 24 $\mu$m.
The RV and proper motion are consistent with membership. The weak excess may
arise from the ISM nebulosity affecting the photometry.

\textit{ID 2} shows only marginal excess at 8 $\mu$m but a significant level in the NIR
and at 24 $\mu$m. It is a highly probable member according to its proper motion.
We resolved it into two objects 1$\arcsec$ apart.
The system is entered in the Washington Visual Double Star Catalog \citep{Worley96}
with only three observations from 1929-1949, leaving
open the question of the physical association of the objects.
Our spectrum of the primary is consistent with late A7-A9 reported in the literature,
and the radial velocity is close to the cluster mean.
The secondary has stronger Ca II lines, lacks Pa lines and
a radial velocity larger by 23 km/s. We speculate that it may
be a background K giant, which would explain the anomalous $H-K$ color of the system.
Does the 24 $\mu$m excess originate from the primary A type member
or the K giant then?
We are inclined to attribute it to the former since excesses around G-K giants are rare
\citep{Zuckerman95,Jura90,Plets97}, while excesses around A stars of this age are very common \citep{Rieke05}. 

\textit{ID 4} in \citet{Jeffries03} is indicated as a K5 radial velocity member with Li absorption 
and weak chromospheric emission in H$\alpha$ (RX 9).
It is interesting that besides a weak 8 $\mu$m excess it also
shows a robust excess at 4.5 $\mu$m and a marginal one at 3.5 and 5.8 $\mu$m.
We did not obtain a spectrum of it.

\textit{ID 7} is the latest-type member with a
robust detection at 24 $\mu$m, indicating an excess of $\sim$3 mag.
It was the largest excess detected in \citet{Young04}.
\citet{Jeffries05} classified it as an M4.5 radial velocity member with no Li.
It does not show excess at 5.8 $\mu$m, but has the reddest $K-[8]$ color
among late-type members (we do not count ID 12 as a reliable detection).
The strong K I at 7699\AA\, and Na I doublet at 8200\AA\, in our spectrum firmly rule out
the possibility of a low-gravity background giant \citep{Torres93, Schiavon97},
while our radial velocity again confirms membership. The spectrum shows emission in H$\delta$
and H$\gamma$, but higher resolution is needed to tell whether the emission is
due to a chromosphere or to accretion.
The 24 $\mu$m excess of ID 7 is surpassed by only one other candidate M-type member,
2MASS 08090344-4859210 (Fig. \ref{fig10}).
With F$_{24}=$1.2 mJy, this source is a clear detection
despite the expected low mass. 
It makes a close match with ID 7 in optical colors, being only 0.15 mag bluer in
$R-I$ and $V-K$, indicating a spectral type of $\sim$M3.
Similarly to ID 7, the excess vanishes in the shorter IRAC bands.
If this object is a sibling of ID 7,
it would mean that stars of $\sim$30 Myr age and $\sim$0.25 M$_{\odot}$
are capable of generating dusty disks
even more vigorously than massive stars.
There is a caveat though. \citet{Jeffries04} dismiss it as a member
based on the $RIZ$ photometry. However, it is only
0.05 mag off the nearest members on the $I$ vs $R-I$ and
$I-Z$ vs $R-I$ diagrams, and is a legitimate member on
$I$ vs $I-Z$, $V$ vs $V-K$ ($R-K$, $I-K$) color-magnitude diagrams.
With $J-H=0.76\pm0.06$, $H-K=0.25\pm0.06$
it is slightly above other late type members on the $J-H$ vs $H-K$ diagram,
with errorbars allowing either a dwarf or a giant interpretation.
Spectroscopy of gravity-sensitive features similar to that for ID 7
is needed to resolve this interesting case.

\textit{ID 8} is a single mid-G member according to the optical and
NIR color-magnitude diagrams.
Our spectrum is consistent with this classification
and indicates that its radial velocity is also consistent with membership.
Together with ID 9, it is unique in that its excess starts already at 3 $\mu$m.

\textit{ID 9} has the biggest excesses at all IRAC wavelengths
and at 24 $\mu$m in our survey.
It was detected already by {\it IRAS}; the Serendipitous Survey Catalog \citep{Kleinmann86}
contains a measurement at 25 $\mu$m and upper limits at 12 and 60 $\mu$m.
The {\it IRAS} flux at 25 $\mu$m is in excellent agreement with ours:
0.16 Jy.
According to the $VRK$ color-magnitude diagrams,
it could be an equal-mass binary late-K dwarf.
Our spectrum is consistent with this classification.
The radial velocity of $+29$ km/s 
is consistent with membership, being within the HWHM from the cluster
value in its radial velocity group (Fig. \ref{fig5}).
The only notable difference between it and the
other members with similar colors
is the shallower and broader lines, indicating
faster than typical rotation with $v\,sin\,i > $20 km/s (see Fig. 4  
in  \citet{Jeffries00}).
Its $H-K$ color of 0.3 mag is however more characteristic of a mid-M dwarf.
We can rule out the latter classification with our spectrum.
It appears therefore that ID 9 has an IR excess starting
at 2 $\mu$m and rising all the way to 24 $\mu$m.

The fraction of 8 $\mu$m excesses for all sources in Tables \ref{table1a} and \ref{table1b}
except for those with peculiarities discussed in footnotes is
$<$5\% (0/20), $\leq$7\% (0 -- 2 out of 28, depending on the marginal cases of
ID1 and ID2), 1.5\% (1/68), 1.0\% (2/208), and 0.4\% (1/227)
for BA, F, G, K and M stars respectively.
When discarding proper motion non-members and applying the proper motion
membership fraction found in \S \ref{pm} to those without proper motion estimates,
we obtain correspondingly
$<$5\% (0/19), $\leq$9\% (0-2/23), 2.6\% (1/(35+7$\times$0.52)), 1.2\% (2/(143+30$\times$0.77)),
$\leq$0.6\%
($\leq$1/(8+216$\times$0.80\footnote{Though high membership probability for M stars in NGC2547 is based
on only 21 stars with known proper motions, it was confirmed
in the RV and Li study of \citet{Jeffries05} with three times larger sample.}),
depending on whether ID7 is a proper motion 
member or not).
Summarizing, we find the 8 $\mu$m excess fraction at 30 Myr
to be $\sim$1\% for the intermediate-mass to low-mass stars.
Among four detected 8 $\mu$m excesses two
(ID 4, ID 7) are weak and two (ID 8, 9) are highly significant;
three of them also have detected 24 $\mu$m excesses.

\subsection{24 $\mu$m excesses}

Excesses at 24 $\mu$m are far more common than at 8 $\mu$m. 
There are 27 objects in Table \ref{tableexc} that show excess
at the former but not at the latter wavelength. 
To compare NGC 2547 to other (mostly older) clusters, we compute the fraction
of members with 24 $\mu$m excess.
As we have seen, by $(V-K)_0 \sim 1.4$ we start
missing members in the MIPS band, while the identified excesses
are prone to confusion with unrelated sources and nebulosity.
We therefore calculate the excess fractions below this limit.
In the range $-0.2 \leq (V-K)_0 < 0.7$, corresponding to
$\sim$B8--A9, 8 out of 18 highly probable members with good photometry show excess,
resulting in an excess fraction of 44\%. Adding 2 possible members of which one
has excess (triangles on Fig.\ref{fig10}) results in 45\% (9 out of 20).
All these sources are also proper motion members,
except for non-excess 2MASS 08094610-4914270,
which however qualifies as member based on the Tycho Catalog measurement\footnote{Available
via the VizieR Online Data Catalog I/250.}.
In the range of $0.7 \leq (V-K)_0 < 1.4$ , F0--F9,
the excess fraction among the highly probable photometric members is 
33\% (7/21) and the same (33\%) with addition of 3 possible members (8/24).
Excluding 4 proper motion non-members, all of which are
non-excess stars, increases the excess fraction in F stars to 40\% (8/20).

\section{Discussion}

\subsection{Evolution of 8 micron excess}

The inner accretion disks
have been extensively studied at 2--4 $\mu$m already from the ground,
and the colder debris disks at 24--70 $\mu$m in the recent large {\it Spitzer} surveys
\citep[e.g.,][]{Kim05, Rieke05, Bryden06, Su06, Meyer07}). Recent {\it Spitzer}
studies are also starting to
put together a picture of the evolution of inner disks
probed at 8 $\mu$m.

Emission at 8 $\mu$m traces dust hotter than $\sim$300 K,
corresponding to distances of $\sim$1--10 AU around G--A stars,
and $\sim$0.1--1 AU around M--K stars.
Non-detection of 8 $\mu$m excess in the majority of
NGC 2547 members signals a thorough cleaning of hot dust by 30 Myr.
Assuming that grains emitting at 8 $\mu$m are astronomical silicates with a
dust temperature of 350 K, grain density of 2.5 g cm$^{-3}$ and a
radius of 10 $\mu$m, a 8 $\mu$m excess of 0.1 mag 
around A--F type stars corresponds to a dust mass of $\sim10^{-4}$M$_{Moon}$, and
$\sim10^{-5}$M$_{Moon}$ for an excess of 0.15 mag around G to early-M
stars in NGC 2547.

Excess at 8 $\mu$m is generally found in young stars
simultaneously with excesses at shorter IRAC bands and at 24 $\mu$m,
implying presence of optically thick primordial disks.
Stars with colors close to photospheric in the shorter IRAC bands but with excesses starting
at $\sim$8--10 $\mu$m are less frequent.
In stars several Myr old such disks
are called ``transition'', meaning they can be in the
evolutionary state from primordial to debris \citep[e.g.,][]{Hartmann05a, Lada06}. 
The reduced emission in the IRAC bands
is interpreted as either grain growth/settling
or clearing by the dynamical effects of giant planets.
Depending on the excess at longer wavelengths, 
further subdivision of transition disks can be found in the literature.
It is also possible that some of the transition-like SEDs 
may actually belong to young debris systems \citep{Currie07, Furlan07, Hernandez07, Rhee07}. 

\citet{Hernandez06} surveyed 60 B--F stars in
the two Orion subassociations OB1b (5 Myr old) and OB1a (10 Myr old).
They found two 8 $\mu$m excesses in the younger group
(one arising from the accretion disk of a Herbig Ae star
and another from a massive debris),
and one excess in the older group (from a Herbig Ae star),
resulting in 8 $\mu$m excess fractions among hot stars of 9\% and 5\%
at 5 and 10 Myr respectively.  
\citet{Sicilia06} studied two clusters in Cepheus OB2 association.
Among 21 B--F stars with 8 $\mu$m measurements in the 4 Myr old Tr 37,
two show excesses, only one of which may be due to a dusty disk
(the other star is of the classical Be type). In the 10 Myr old NGC 7160
2 out of 68 stars show weak 8 $\mu$m excesses.
These fractions, 5--10\% at 4 Myr and 3\% at 10 Myr are quite
consistent with the Orion results, indicating a fast decline of 8 $\mu$m emission
in massive stars.      
The fraction of G--M stars with measured 8 $\mu$m excess
is much higher: $\sim$60\% (of which 15\% are classified
as transitional) in Tr 37 and 5--8\% in NGC 7160
(1--2 of excesses are transitional and the other one is primordial).
  
\citet{Megeath05} studied 15 K--M stars in the $\eta$ Chameleon 5--9 Myr association.
They found 6 to have excesses at the IRAC bands (excess fraction 40\%),
five of these are accretors, one not.
None of the two A members shows excess.
\citet{Low05} summarize SEDs for the 8--10 Myr old TW Hydrae
association. Among 22 K--M members, 3 show excesses
at 8--10 $\mu$m, two of which are from primordial disks (TW Hya and Hen 3-600)
and one from a debris disk (HD 98800AB), resulting in an excess fraction of 14\%.
\citet{Curry06} found in the 13 Myr-old double cluster $\chi$ \& $h$ Persei
8 $\mu$m excesses in 1--3\% of the intermediate-mass stars
and up to 4--8\% in the solar-mass stars (at their completeness
limit of 1.2--1.4 M$_{\odot}$). 
\citet{Silverstone06} surveyed 74 solar-mass (G--mid-K) young nearby stars
with IRAC and MIPS. They found 5 stars with IRAC excesses,
all of them from optically-thick disks;
all are younger than 20 Myr. The
excess fractions are 14$^{+11}_{- 7}$ \%  and 2$^{+5}_{-2}$ \%
for the 3--10 Myr and 10--30 Myr bins respectively.

Although there is some scatter (probably in part due to statistical
uncertainties in relatively small samples), 
these studies agree on a dramatic overall decay in 8 $\mu$m excess incidence
over the first 20 Myr of stellar evolution. Beyond this age such
excesses become very uncommon, and based on the dynamic arguments of the grain
survival time are expected to have debris origin.
At 30 Myr among $\sim$500 B--M stars in NGC 2547 we identify
4 with IRAC excesses, two of which are only marginal,
while the other two are extremely strong (\S \ref{trans}).
\citet{Mamajek04} studied 14 F--M members of the 30 Myr old Tucana-Horologium association
and found none with 10 $\mu$m excess, resulting in an excess fraction $<$7\%.
The less homogeneous {\it Spitzer} studies of field stars have yielded only a few
discoveries of hot disks beyond 30 Myr.
\citet{Chen06} obtained {\it Spitzer} IRS spectra for 59 nearby mostly B--F stars with {\it IRAS}
60 $\mu$m excesses.
They found 7 at ages $\leq$10 Myr, 2 at ages 20--50 Myr and
2 at 500-1000 Myr ($\zeta$ Lep and $\eta$ Crv)
to have 8-13 $\mu$m excess (in amount $\geqslant$10\% above the photospheric flux).
From a similar study of 41 main sequence F5--K5 stars \citet{Beichman06}
found another two (HD 69830 and HD 72905), only one of which is older
than 1 Gyr, resulting in an excess fraction of 2.5\% among old stars.
\citet{Uzpen05,Uzpen06} searched the MSX and the GLIMPSE {\it Spitzer} Legacy Program fields
in the Galactic plane for excess due to hot circumstellar disks.
They found $\sim$50 such candidates among B--K field stars.
Unfortunately we can not include them
in our analysis due to the lack of information about
their ages, but a significant proportion show signs of youth \citep{Uzpen06}.

As we have seen, unlike the shorter IRAC wavebands,
the 8 $\mu$m excess can trace both primordial and debris material,
and therefore should be used in combination with other diagnostics
to interpret its overall decay with age.
The above studies indicate an inside-out evolution for \textit{primordial} disks.
The inner-most parts of the disks (within few stellar radii),
as traced by the NIR photometry and spectroscopic signatures of accretion,
clear out on a time scale of $\sim$3 Myr, with the disk fraction falling by 10 Myr to a few percent
for solar-mass stars and essentially to zero
for intermediate-mass ones. Excess at 6--8 $\mu$m holds longer,
being at a level of 5--10\% for low-mass stars
at 10 Myr and dropping to 1\% by 30 Myr. 
The evolution of \textit{debris} disks is more slowly,
with a characteristic time scale of 150 Myr for the 24 $\mu$m
excess \citep{Rieke05, Gorlova06, Siegler06, Su06}
and $\sim$400 Myr for 70 $\mu$m \citep{Su06}.
The collisional cascades beyond 10 Myr are maintained
with the formation of the large icy bodies in the outer disk
that are capable to steer leftover planetesimals
\citep{Dominik03, Kenyon04b, Kenyon05, Hernandez06}.
Unlike 24 $\mu$m excess, 8 $\mu$m excess does not show a secondary peak after first few Myrs,
and is less frequent among B--A stars.
This behavior may imply a rapid depletion of the material within 1 AU,
inhibiting planet formation and the corresponding debris stage in this radial zone.
Even if planets do form in a few Myr,
as rare cases with inner holes suggest,
the end process of planet formation in this zone must lead to 
the rapid dissipation of smaller bodies to inhibit
collisional cascades (e.g., due to the absence of icy bodies
or because of the rapid inward planet migration
\citep{Burkert06}).

\subsection{Transitional disks at 30 Myr?}\label{trans}

\begin{figure} 
\epsscale{0.9}
\plotone{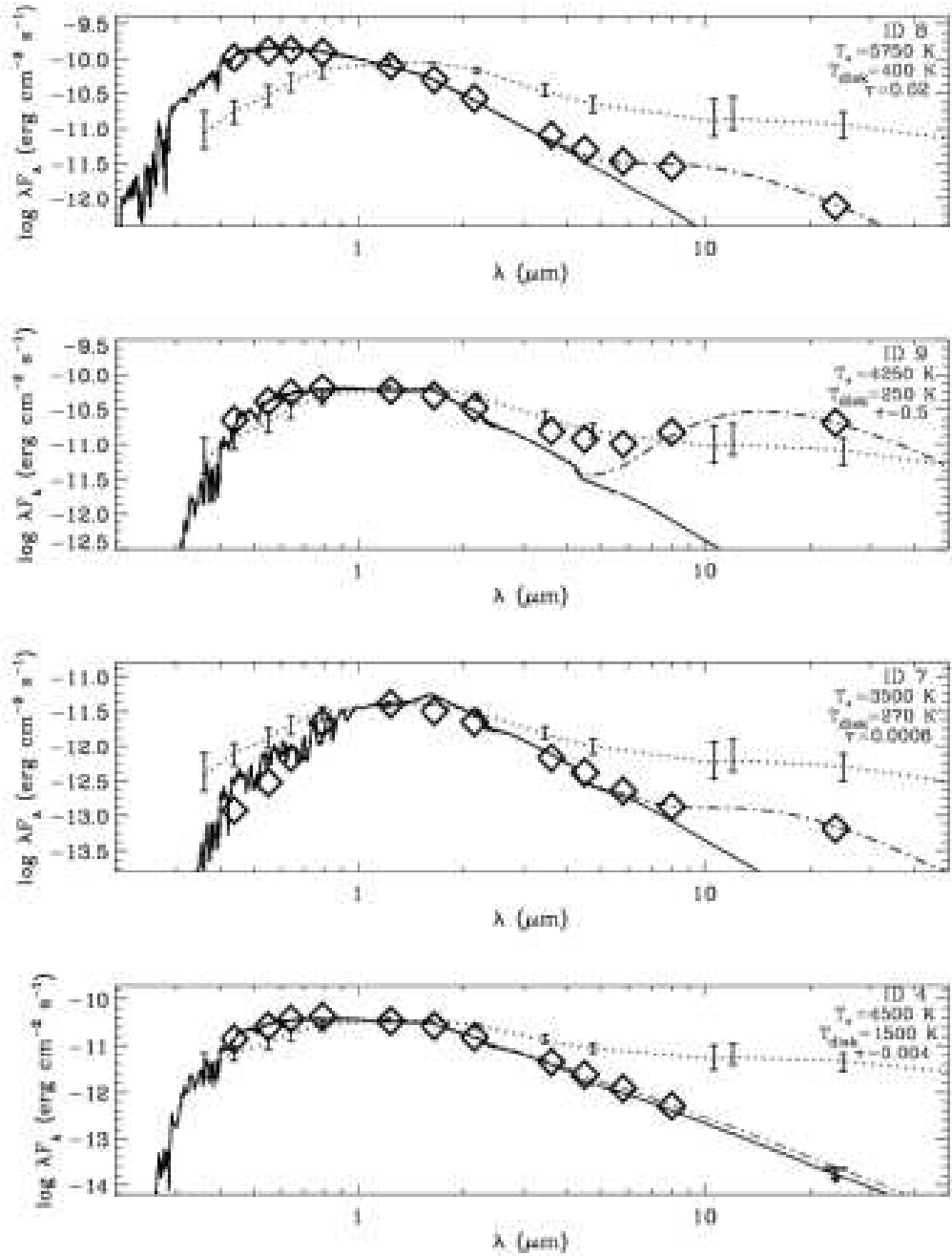}
\caption{Spectral energy distributions of stars with 8 $\mu$m excess.
{\it Diamonds:} $BVRIJHK$, IRAC and MIPS 24 $\mu$m derddened photometry
($A_{V}=0.19$, extinction coefficients from \citet{Mathis90});
{\it solid line:} Kurucz photospheric model spectra with
effective temperature T$_{*}$ and $log\,g=3.0$; {\it dot-dashed line:}  photospheric
flux plus a black-body flux of temperature T$_{disk}$ and fractional luminosity
$\tau$; {\it dotted line:} median SED with the quartiles of the distribution
for the class II K7--M2 sources in the Taurus. 
}\label{fig12}   
\end{figure}

De-reddened spectral energy distributions of NGC 2547
sources with 8 $\mu$m excess are presented in Fig. \ref{fig12}.
First we over-plot Kurucz models with effective temperatures corresponding to
the expected spectral types and sub-giant gravity,
normalizing them to fit the observations in the J band.
The flux beyond 10 $\mu$m was extrapolated with the Planck function.
As one can see, photospheric models match 
the SEDs of the excess stars well from 0.4 (B band) to 2 $\mu$m (K),
with only a slightly worse fit for ID 7 (M4.5) with strong molecular features.
These fits confirm that excess emission is present in all \textit{Spitzer} bands in
ID 8 and 9, and starts at 8 $\mu$m in ID 7, a result we obtained
previously from empirical color-color diagrams.
We fit the excess emission with a series of black bodies,
by varying the temperature and the normalization factor.
The best fits together with the fraction of the excess flux
relative to the photospheric value $\tau$ ($=L_{Exc}/L_{*Bol}$)
are given in Fig. \ref{fig12}. For comparison, we also show
the median SED for the low-mass stars with optically-thick disks in the 1--2 Myr-old Taurus region
\citep{Dalessio99,Hartmann05a}.    
We could not obtain a satisfactory single-temperature fit for ID 9,
so a 250 K black body curve is shown for reference only. 
For ID 4 the small IRAC excesses and the low upper limit on 24 $\mu$m flux
require temperature for excess emission of no less than 1000 K.
The best fit is achieved with a maximum possible temperature of 1500 K,
corresponding to dust sublimation temperature.
While the fractional excess luminosities of ID 4 and 7 place them
in the range of typical debris systems, the luminosities of ID 8 and 9
are too big, suggesting that at least ID 9 must have an optically-thick disk. 

Three possibilities exist to explain the excesses of ID 8 and 9:
1) the disks are transitional, evolving
from primordial to debris (as seen in the smaller IRAC excesses
compared to Taurus),
2) a catastrophic collision occurred between two planetary-size bodies,
3) these objects are background post-MS stars experiencing mass-loss.

The 3.5--5.8 $\mu$m colors of both ID 8 and 9 are consistent with colors of
classical T Tau stars (CTTS) \citep{Hartmann05a}.
The 8 and 24 $\mu$m excesses of ID 8
are also more characteristic of CTTS and only rarely observed in weak T Tau stars,
while the excesses of ID 9 are extreme even for CTTS \citep{Padgett06}.
Despite big mid-IR excesses, the two stars however
lack strong excess in the NIR (ID 8 has none).
They also lack emission for example in H$\delta$ and the Ca II triplet in our spectra.
These are characteristics of transitional disks
where gas accretion has ceased and dust has dissipated
from the inner 0.1 AU,
but primordial dust at larger radii is still present.
The question is whether the 30 Myr age of the cluster 
is widely discrepant with this interpretation.
The typical ages of T Tau and transitional disks are a few Myr.
On the other hand, K-band excess may persist at the level of few percent
at ages of 10 Myr \citep{Hillenbrand05,Bonatto06}.
The cases of gas accretion beyond 10 Myr are rare
but also exist, the most famous being a low-mass Classical T Tau star St 34
in the Taurus, whose Li age is at least 25 Myr according to \citet{White05} and \citet{Hartmann05b}. 
They suggest that the long-lived accretion is maintained
due to binarity of this star with sub-AU separation.
ID 8 does not appear to be a binary though. 
The nearby associations like TW Hya and $\eta$ Cha present the best
examples of how diverse disks can be at 10 Myr \citep{Low05,Riaz06,Megeath05}. 
Considering that our study is one of the first that explore
disks at 0.1--1 AU at the 20--40 Myr age range with good statistics, 
we can not dismiss the possibility that ID 8 and 9 may
be the oldest known transitional disks around solar-mass stars.

Interesting, NGC 2547 is at a similar era when the Earth-Moon system
formed. It is believed that the Earth-Moon system resulted from a giant
impact between the proto-Earth and a Mars-sized body \citep{canup04}.
After the impact, it is thought that the proto-Earth accreted less
than 5\% of a current Earth mass from the impactor, while $\sim$30\%
of the impactor mass was vaporized and escaped. Only 1.5 to 2 M$_{Moon}$
of material stayed bound to the proto-Earth orbit, forming a disk around
the Earth that eventually consolidated into the Moon.
We have explored whether such a giant impact might be
responsible for the infrared excess we observe around ID 8.
Using an optically thin disk model with the
assumptions of constant surface density and a power-law grain size
distribution ($\Delta N\sim a^{-3.5}\Delta a$) of astronomical silicates, the 8 and 24 $\mu$m excesses
are consistent with a disk with an inner radius of $\sim$0.1 AU (dust
sublimation radius for 0.1 $\mu$m grains) and an outer radius of
$\sim$1 AU. A wide range of grain size distributions can yield
reasonable fits to the data, but all the fits require a dust mass on
the order of 10$^{-3}$ M$_{Moon}$. The mass we derived here is a lower
mass limit of the total material around ID 8 since we used an
optically thin model. In addition, the mid-infrared observations do not
trace the population of large (km-size) parent
bodies. Following the same approach as for HD 69830 by
\citet{beichman05}, we estimate a total mass of parent bodies on the
order of 3-10 M$_{Moon}$, assuming the maximum size of the parent
bodies is 1 to 10 km. 

\begin{figure} 
\plotone{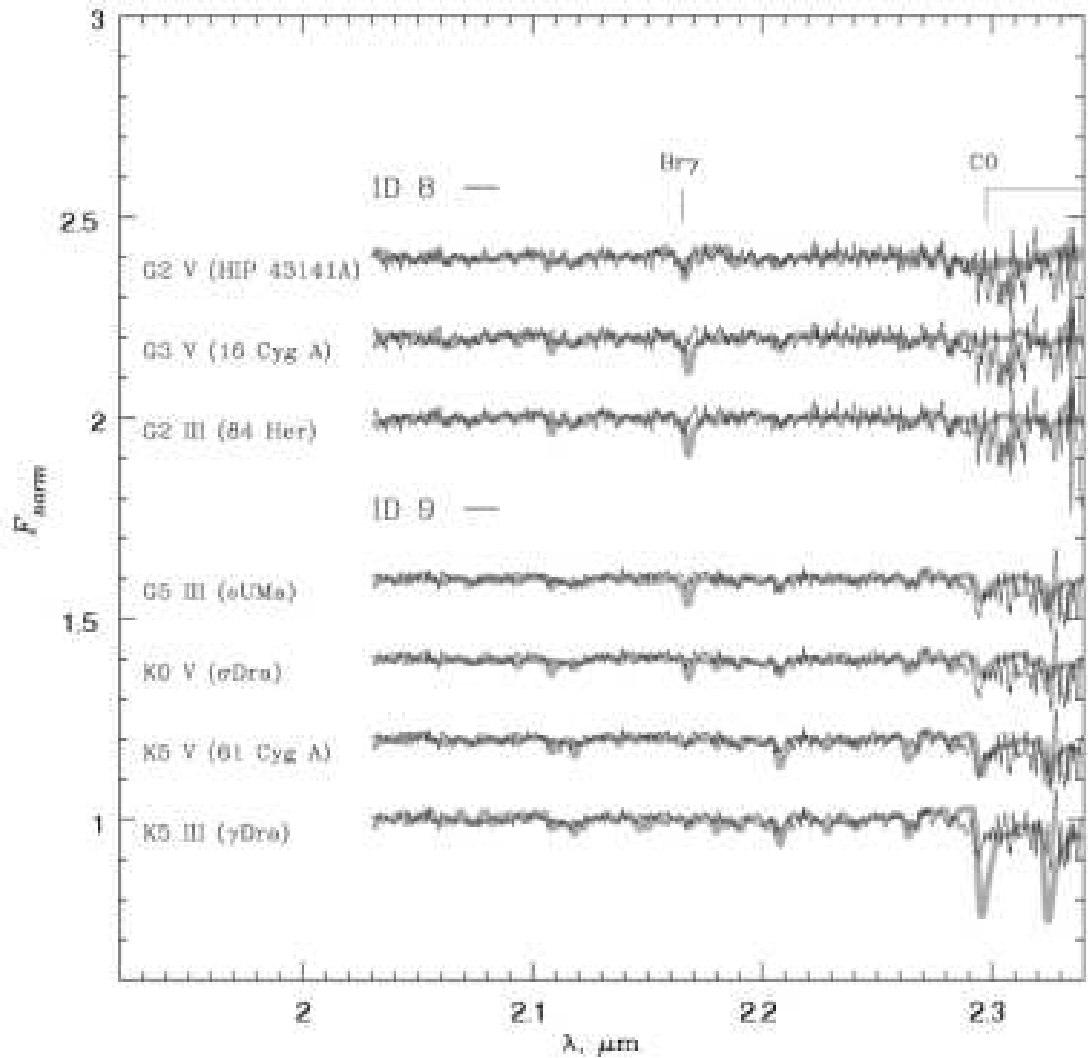}
\caption{SofI spectra of ID 8 \& 9 (\textit{thin black lines}) compared to the
spectra of standard stars from \citet{KH86}, and to HIP 43141A obtained on the same night
(\textit{thick gray lines}). Standard spectra have been smoothed
to match the SofI resolution. All spectra have been continuum normalized
and shifted in the vertical direction  (except $\gamma$ Dra).
The spectrum of ID 8 is consistent with a G type,
while the spectrum of ID 9 matches a mid-K dwarf. 
}\label{fig13}   
\end{figure}

Could ID 8 \& 9 be interpreted as
background interlopers with dusty envelopes?
Indeed, ID 9 is in the region of the cluster color-magnitude
sequence that is most strongly contaminated by Galactic G--K giants
\citep{Littlefair03}. ID 8 is not, but it comfortably lies
among dusty evolved stars on the $[3.5]-[8]$ vs. $[8]-[24]$
diagram of the LMC field in \citet{Meixner06}.
Despite their big mid-IR excess, the $H-K$ excess in ID 8 and 9
is much weaker compared with the AGB stars \citep{Buchanan06}.
Perhaps ID 8 and 9 then belong to a rare class of the
first-ascent G--K giants where IR excess originates
from a shell thrown off as a result of a mixing event
in the stellar interior \citep{deLaReza96,Drake05}?
CO first-overtone bands at 2.3 $\mu$m can be used to probe
the background giant explanation for ID 8 and 9
from low-resolution spectra.
This feature appears at spectral type mid-G and strengthens
at lower temperatures. It is also
strongly gravity-sensitive, for the same spectral type
being deeper in supergiants and giants ($log\,g \leq 3$)
than in dwarfs ($log\,g \sim 4.5$) \citep{KH86,Wallace97,Ivanov04}. 
According to the evolutionary models
\citep{Siess00}, G-K stars in NGC 2547 are close enough to the Main Sequence
to have dwarf gravities within a few tenths of a dex;
therefore, the CO band should be weak in them.
In Fig. \ref{fig13} we show our K-band spectra of ID 8 \& 9
overplotted on the spectra of field stars from \citet{KH86}\footnote{Available
in the electronic form from the VizieR on-line catalog III/114},
and a spectrum of HIP 43141A obtained
and reduced by us in the same manner as ID 8 \& 9.
The spectra of ID 8 \& 9
lack strong features, consistent
with G to mid-K spectral types expected from observed colors
and negligible extinction in the cluster.
ID 8 shows a weak but distinctive Br$\gamma$ line,
indicating an earlier SpT than ID 9.
ID 9 in turn clearly shows a CO(2,0) band-head at 2.294$\mu$m,
which best compares with the K5 dwarf 61 Cyg A.
Despite residual telluric noise in that region,
we can firmly rule out any possibility for ID 8 \& 9
to be common late K -- M giants.  
However at this low resolution
we can not exclude possibility
of ID 8 \& 9 being mass-losing G supergiants,
with simultaneously shallow Br$\gamma$ and CO lines
due to filling emission and residuals
induced in the telluric correction process.
Such stars are however extremely rare \citep{Oudmaijer95}.
In summary, the near-IR spectra of ID 8 \& 9 are
consistent with a mid-G and mid-K dwarf types respectively,
supporting their membership in the cluster. 

\section{Conclusions}

We have extended the survey of \citet{Young04}
of the circumstellar disks in the 30 Myr old cluster NGC 2547
to include the newly identified members over the
entire $1.5\degr \times 1\degr$ area covered by {\it Spitzer}.
Among $\sim$600 B--mid-M probable members (corresponding to masses down to 0.2 M$_{\odot}$)
where we are sensitive to the photospheric level,
only 4 stars show convincing excesses in the IRAC bands,
corresponding to an excess fraction $\lesssim$1\%.
This behavior is in striking contrast with the 24 $\mu$m excesses
exhibited by 30--45\% of the B--F members.

Of the four 8 $\mu$m excesses, two are marginal
and have SEDs consistent with debris disks (around an M4 dwarf discovered
by \citet{Young04} and a K5 dwarf).
The other two members have strong 8 $\mu$m excesses.
One is a mid-G dwarf with SED consistent with a massive
hot debris disk, perhaps the result of a giant impact.
Another is a mid-K dwarf whose SED
is more consistent with an optically-thick disk,
though with depressed 2--6 $\mu$m emission compared
with the average SED of a CTTS.
Though we do not find signatures of gas accretion in
our spectra of these two stars, spectroscopic
investigation at higher resolution is needed
to confirm the transitional nature of
these disks.
We identify another possible mid-M member with excess
starting between 8--24 $\mu$m; if confirmed, it will be
one of the few known M dwarfs older than 20 Myr with mid-IR excess
\citep{Siegler06}.

Our survey provides the best evidence to date that 
the frequency of the 8 $\mu$m excess that traces the disk zone inside $\sim$1 AU
around solar-type stars diminishes to less than $\sim$1\%
by 30 Myr. If planetary formation occurs inside 1 AU, it must be complete by 30 Myrs.
\\
\\
\textbf{Note in proof.} The recent paper by \citet{Naylor06}
obtains  a somewhat older mean age of 38.5$^{+3.5}_{-6.5}$ Myr for NGC 2547
(for a smaller distance of 361$^{+19}_{-8}$ pc). This estimate of the age
is within the errors of the previous estimates.
The conclusions of our paper are not affected significantly if
we adopt the new age estimate.

\acknowledgments 

We would like to thank J. Harris, D. Kelson, A. Oemler, G. Walth and Y. Yang 
for help with IMACS reduction, and J. S. Kim and N. Siegler for help with
observations. An anonymous referee provided many insightful comments.
This paper includes data gathered with the 6.5 meter 
Magellan Telescopes located at Las Campanas Observatory, Chile.
This publication makes use of data products from the Two Micron All Sky Survey,
which is a joint project of the University of Massachusetts 
and the Infrared Processing and Analysis Center/California Institute 
of Technology, funded by the National Aeronautics and Space Administration 
and the National Science Foundation.
This work was supported by JPL/Caltech under contract 1255094.

\clearpage
\begin{deluxetable}{llrrrrrrrrrrrrrrrrl} 
\tablecolumns{19}  
\tablewidth{0pt} 
\tabletypesize{\tiny} 
\tablecaption{\textit{Spitzer} photometry for NGC 2547 highly-probable photometric members\label{table1a}}
\tablehead{ \colhead{RA $\degr$} & \colhead{DEC $\degr$} & \textit{V-K} & \textit{R$_{C}$-K} & \textit{K}\tablenotemark{1} &
 \it{$\pm$} & \it{[3.6]} & \it{$\pm$} & \it{[4.5]} & \it{$\pm$} & \it{[5.8]}  & \it{$\pm$} & \it{[8.0]} & \it{$\pm$} &
 \it{[24]}  & \it{$\pm$} &  Field & Star &  \\ 
\multicolumn{2}{c}{(2000, 2MASS)} &  &  &  &  &  &  &  & & & & & & & &  No.\tablenotemark{2} & No.\tablenotemark{2} & 
} 
\startdata 
122.532396 & -48.919441 &  99.99 &   1.17 &  11.62 &   0.02 &  11.61 &   0.00 &  11.63 &   0.01 &  11.66 &   0.02 &  11.56 &   0.03 &  99.99 &  99.99  &  23 &    27  &  Y \\
122.532740 & -49.276600 &  -0.01 &  99.99 &   8.43 &   0.02 &   8.50 &   0.01 &   8.46 &   0.01 &   8.49 &   0.00 &   8.51 &   0.00 &   7.64\tablenotemark{h} &   0.04 &    9 &  8432  &  Y \\
122.534364 & -49.183403 &  99.99 &  99.99 &  99.99 &  99.99 &  14.99 &   0.02 &  15.01 &   0.05 &  99.99 &  99.99 &  99.99 &  99.99 &  99.99 &  99.99 &    13 &  1776  &  N/A \\
122.534440 & -49.332378 &  99.99 &   1.16 &  11.21 &   0.02 &  11.18 &   0.00 &  11.22 &   0.00 &  11.23 &   0.01 &  11.16 &   0.01 &  99.99 &  99.99 &    8 &    15  &  Y \\
122.535073 & -49.012066 &   0.29 &  99.99 &   9.55 &   0.02 &   9.54 &   0.01 &   9.64 &   0.00 &   9.54 &   0.01 &   9.54 &   0.00 &   8.43 &   0.02 &    30 &  8420  &  Y \\
122.537235 & -49.261490 &   3.28 &   2.58 &  12.19 &   0.03 &  12.06 &   0.00 &  12.13 &   0.01 &  12.06 &   0.03 &  12.03 &   0.08 &  99.99 &  99.99 &    13 &   174  &  N \\
122.538460 & -49.048122 &  99.99 &   2.17 &  12.27 &   0.03 &  12.24 &   0.01 &  12.34 &   0.01 &  12.22 &   0.02 &  12.25 &   0.04 &  99.99 &  99.99 &    18 &   103  &  Y \\
122.538682 & -49.348309 &   1.64 &   1.27 &  10.81 &   0.02 &  10.82 &   0.00 &  10.82 &   0.00 &  10.88 &   0.01 &  10.81 &   0.01 &  10.74 &   0.16 &    8 &    19  &  Y \\
122.539099 & -49.015720 &   2.06 &  99.99 &  10.43 &   0.02 &  10.42 &   0.00 &  10.59 &   0.00 &  10.37 &   0.01 &  10.36 &   0.01 &  10.35 &   0.11 &   18 &    43  &  Y \\
122.539829 & -48.942863 &  99.99 &   2.54 &  12.22 &   0.03 &  12.14 &   0.00 &  12.23 &   0.01 &  12.21 &   0.02 &  12.14 &   0.04 &  99.99 &  99.99 &   18 &   116  &  Y \\
122.540051 & -49.265015 &   2.89 &   2.24 &  11.20 &   0.02 &  11.13 &   0.00 &  11.20 &   0.00 &  11.11 &   0.01 &  10.97\tablenotemark{c} &   0.03 &  10.57 &   0.14 &    13 &   114  & Y \\
122.541465 & -49.609333 &  99.99 &   2.67 &  11.33 &   0.03 &  11.21 &   0.00 &  11.42 &   0.00 &  11.31 &   0.01 &  11.28 &   0.02 &  99.99 &  99.99 &     3 &    32  &  Y \\
\\
           &            &        &        &        &        &        &        &        &        &        &        &        &        &        &         &  PM?\tablenotemark{3}    &     &    \\
122.532396 & -48.919441 &        &        &        &        &        &        &        &        &        &        &        &        &        &         &   Y  &       &    \\
122.532740 & -49.276600 &        &        &        &        &        &        &        &        &        &        &        &        &                         &        & Y     &       &    \\
122.534364 & -49.183403 &        &        &   &  &   &    &  &        &        &        &        &        &        &        &   N/A    &     &      \\
122.534440 & -49.332378 &        &        &  &   &        &        &        &        &        &        &        &        &        &        &  Y    &       &    \\
122.535073 & -49.012066 &        &        &   &  &        &        &        &        &        &        &        &        &        &        &    Y   &       &    \\
122.537235 & -49.261490 &        &        &   &        &        &        &        &        &        &        &        &        &        &        &    N   &       &    \\
122.538460 & -49.048122 &        &        &   &        &        &        &        &        &        &        &        &        &        &        &     Y  &       &    \\
122.538682 & -49.348309 &        &        &   &        &        &        &        &        &        &        &        &        &        &        &    Y  &       &    \\
122.539099 & -49.015720 &        &        &  &        &        &        &        &        &        &        &        &        &        &        &    Y  &      &    \\
122.539829 & -48.942863 &        &        &  &        &        &        &        &        &        &        &        &        &        &        &    Y  &       &    \\
122.540051 & -49.265015 &        &        &   &        &        &        &        &        &        &        &                         &        &        &        & Y      &       &   \\
122.541465 & -49.609333 &        &        &   &        &        &        &        &        &        &        &        &        &        &        &     Y  &       &    \\
\enddata 
\tablenotetext{1}{\textit{K$_{S}$} from 2MASS} 
\tablenotetext{2}{Designations from  ``Wide catalog'' of \citet{Naylor02} or from ``$RIZ$'' catalog of \citet{Jeffries04}}
\tablenotetext{3}{Proper motion membership flag (\S \ref{memb}), N/A: proper motion not available}
\tablenotetext{h}{Extended (\underline{h}alo-like) sources at 24 $\mu$m} 
\tablenotetext{c}{Excess at 8 $\mu$m may be due to \underline{c}ontamination of the aperture by
a nearby source, cosmic ray or strong nebulosity}
\tablecomments{
\\
Objects from this table are shown as solid circles
on Figures \ref{fig6}, \ref{fig7}, \ref{fig8}, \ref{fig9}, \ref{fig10}, \ref{fig11},
except for those with questionable photometry (notes h, c).\\
99.99 means no data.
The complete version of this table will be in the electronic edition of
the ApJ. The printed edition contains only a sample.} 
\end{deluxetable}  
 
\clearpage
\begin{deluxetable}{llrrrrrrrrrrrrrrrrl} 
\tablecolumns{19}  
\tablewidth{0pt} 
\tabletypesize{\tiny} 
\tablecaption{\textit{Spitzer} photometry for NGC 2547 less plausible photometric members\label{table1b}}
\tablehead{ \colhead{RA $\degr$} & \colhead{DEC $\degr$} & \textit{V-K} & \textit{R$_{C}$-K} & \textit{K} &
 \it{$\pm$} & \it{[3.6]} & \it{$\pm$} & \it{[4.5]} & \it{$\pm$} & \it{[5.8]}  & \it{$\pm$} & \it{[8.0]} & \it{$\pm$} &
 \it{[24]}  & \it{$\pm$} &  Field & Star &   \\ 
\multicolumn{2}{c}{(2000, 2MASS)} &  &  &  &  &  &  &  & & & & & & & &  No. & No. & 
} 
\startdata 
122.716280 & -48.971233 &   2.84 &  99.99 &  11.06 &   0.02 &  11.02 &   0.00 &  11.14 &   0.00 &  11.08 &   0.01 &  10.98 &   0.01 &  99.99 &  99.99 &    18 &   490 &  Y \\
122.730125 & -49.443478 &   4.84 &  99.99 &  13.80 &   0.04 &  13.53 &   0.01 &  13.47 &   0.01 &  13.42 &   0.06 &  13.45 &   0.08 &  99.99 &  99.99 &     9 &  1394 &  N/A \\
122.747649 & -49.142479 &   5.12 &  99.99 &  13.50 &   0.04 &  13.23 &   0.01 &  13.19 &   0.01 &  13.19 &   0.06 &  13.18 &   0.08 &  99.99 &  99.99 &    12 &  1176 &  N/A \\
122.769807 & -49.054371 &   3.96 &  99.99 &  11.85 &   0.04 &  11.73 &   0.01 &  11.64 &   0.01 &  11.60 &   0.02 &  11.64 &   0.04 &  99.99 &  99.99 &    19 &  1086 &  N \\
122.802249 & -49.093540 &   5.60 &  99.99 &  14.43 &   0.07 &  14.12 &   0.04 &  14.09 &   0.04 &  13.72 &   0.11 &  13.41\tablenotemark{c} &   0.10 &  99.99 &  99.99 &    19 &   633  &  N/A \\
122.803718 & -49.162022 &   2.50 &  99.99 &  10.85 &   0.02 &  10.80 &   0.00 &  10.89 &   0.00 &  10.84 &   0.01 &  10.81 &   0.01 &  99.99 &  99.99 &    12 &   137 &  N \\
122.819776 & -49.115688 &  99.99 &   4.02 &  14.80 &   0.10 &  14.47 &   0.02 &  14.36 &   0.03 &  14.22 &   0.10 &  13.94 &   0.14 &  99.99 &  99.99 &    12 &   731\tablenotemark{nm} & N/A \\
122.836424 & -49.219448 &   1.35 &  99.99 &  10.69 &   0.02 &  10.68 &   0.00 &  10.66 &   0.00 &  10.71 &   0.01 &  10.66 &   0.01 &  99.99 &  99.99 &   12 &    22 &  Y \\
122.836981 & -49.151585 &   1.68 &  99.99 &  11.39 &   0.02 &  11.31 &   0.00 &  11.39 &   0.01 &  11.36 &   0.02 &  11.29 &   0.02 &  99.99 &  99.99 &    12 &    64 &  N \\
122.899746 & -48.946228 &   3.17 &  99.99 &  10.95 &   0.02 &  10.90 &   0.00 &  11.00 &   0.00 &  10.96 &   0.01 &  10.88 &   0.01 &  99.99 &  99.99 &    19 &   910 &  Y \\
122.910886 & -49.239845 &   5.29 &  99.99 &  13.87 &   0.05 &  13.70 &   0.01 &  13.69 &   0.02 &  13.58 &   0.06 &  13.68 &   0.09 &  99.99 &  99.99 &   12 &  2353 &  N/A \\
122.945855 & -49.215736 &   3.55 &  99.99 &  12.06 &   0.02 &  11.96 &   0.00 &  12.05 &   0.01 &  11.94 &   0.02 &  12.00 &   0.03 &  99.99 &  99.99 &    12 &  1013 &  N \\
\\
           &            &        &        &        &        &        &        &        &        &        &        &        &        &        &         &  PM?\tablenotemark{3}    &     &    \\
122.716280 & -48.971233 & &  &        &        &        &        &        &        &        &        &        &        & &        &      Y  &    & \\
122.730125 & -49.443478 & &  &        &        &        &        &        &        &        &        &        &        & &        &     N/A   &   & \\
122.747649 & -49.142479 &  & &        &        &        &        &        &        &        &        &        &        & &        &      N/A  &    & \\
122.769807 & -49.054371 &  &  &        &        &        &        &        &        &        &        &        &        & &        &      N  &    & \\
122.802249 & -49.093540 &   &  &        &        &        &        &        &        &        &        &        &        & &        &     N/A   &    &\\
122.803718 & -49.162022 &  & &        &        &        &        &        &        &        &        &        &        & &        &      N  &     & \\ 
122.819776 & -49.115688 &  & &        &        &        &        &        &        &        &        &        &        & &        &     N/A   &   & \\
122.836424 & -49.219448 &  & &        &        &        &        &        &        &        &        &        &        & &        &      Y  &    & \\
122.836981 & -49.151585 & &  &        &        &        &        &        &        &        &        &        &        & &        &     N   &    & \\
122.899746 & -48.946228 & &  &        &        &        &        &        &        &        &        &        &        & &        &     Y   &    &\\
122.910886 & -49.239845 &   &  &        &        &        &        &        &        &        &        &        &        & &        &    N/A    &    &\\
122.945855 & -49.215736 &  &  &        &        &        &        &        &        &        &        &        &        & &        &      N  &    &\\
\enddata 
\tablenotetext{ns}{\underline{N}on-\underline{s}tellar flag in \citet{Jeffries04}}
\tablenotetext{nm}{Below the cluster sequence on the $V-(V-K)$ or $R-(R-K)$ diagram -- \underline{n}on\underline{m}ember?}
\tablenotetext{c}{Excess at 8 $\mu$m may be due to \underline{c}ontamination of the aperture by
a nearby source}
\tablecomments{
\\
Objects from this table are shown as triangles on Fig. \ref{fig6}, \ref{fig8}, \ref{fig10}, 
except for those with notes.\\ 
For description of columns see Table \ref{table1a}.
The complete version of this table will be in the electronic edition of
the ApJ. The printed edition contains only a sample.} 
\end{deluxetable} 

\clearpage
\begin{deluxetable}{ccclccrr} 
\tablecolumns{8}  
\tablewidth{0pt} 
\tabletypesize{\tiny} 
\tablecaption{Log of IMACS observations\label{tablemasks}}
\tablehead{\colhead{Mask No.} & \multicolumn{2}{c}{Mask Center} & \colhead{UT Date} & \colhead{Airmass} & \colhead{Seeing} & \colhead{Integr. Time} & \colhead{$V$} \\
 & RA (h:min:s) & DEC ($\degr:\arcmin:\arcsec$) & & &  FWHM($\arcsec$) & sec $\times$ frames & mag
} 
\startdata
1  & 08:09:40.000 & -49:12:00.00 & 2005 Feb 02 & 1.11--1.12 & 0.6 &  15 $\times$ 7 &   9-13 \\   
2  & 08:09:40.000 & -49:12:00.00 & 2005 Dec 23 & 1.07--1.11 & 0.6 & 900 $\times$ 3 &  13-19 \\   
3  & 08:10:36.000 & -49:15:00.00 & 2005 Dec 21 & 1.09--1.10 & 0.6 & 150 $\times$ 3 &  10-12 \\  
4  & 08:10:36.000 & -49:15:00.00 & 2005 Feb 03 & 1.13--1.18 & 0.7 & 600 $\times$ 3 &  13-15 \\ 
5  & 08:09:40.000 & -48:57:30.00 & 2005 Dec 21 & 1.07--1.07 & 0.6 & 900 $\times$ 3 &  13-18 \\ 
\enddata 
\end{deluxetable}  

\clearpage
\LongTables
\begin{deluxetable}{crrrrccrrrcl} 
\tablecolumns{12}  
\tablewidth{0pt} 
\tabletypesize{\tiny} 
\tablecaption{IMACS spectroscopy of NGC 2547 candidate members\label{tablesp}}
\tablehead{2MASS & $(V-K)_{0}$ & 
$V_{0}$ & $(K-[8])_{0}$ & $(K-[24])_{0}$ &
SpT\tablenotemark{1} & RV\tablenotemark{2} & Prob.\tablenotemark{3}\ & RV\tablenotemark{4} & S/N\tablenotemark{5} & Sp. No\tablenotemark{6} & Remarks\tablenotemark{7} \\ 
 & & & & & & group & \% & km/s & & mask:slit & 
}
\startdata 
 08095753-4908202 & -0.20 & 8.73 & 0.05 & 1.62 & A0Vn & BF   & 37 & 67 & 83    & 1:23  & Broad lines \\
 08093671-4911383 & 0.02 & 8.67 & -0.05 & -0.05 & A0-A1 & BF   & 52 & 59   & 280  & 1:19  & Broad lines \\
 08100607-4914180 & 0.04 & 9.68 & 0.02 & 1.47 & B8-A1 & BF   & 59 & 32   & 38   & 1:18  & Low S/N \\
 08095221-4911022 & 0.06 & 9.48 & 99.99 & 99.99 & 99.99 & BF   & 54 & 34   & 120  & 1:7  & Broad lines \\
 08092958-4906166 & 0.35 & 10.03 & 99.99 & 99.99 & A0-A2V & BF  & 48 & 19   & 180 & 1:22  & Broad lines \\
 08095066-4912493 & 0.36 & 10.03 & -0.05 & 0.02 & A2-A3IV & BF   & 66 & 18   & 93 & 1:21  &  \\
 08095060-4913194 & 0.53 & 10.40 & 99.99 & 99.99 & A0 & G    & 0 & 12   & 95 & 3:7  & Broad lines \\
 08094610-4914270 & 0.53 & 10.49 & -0.00 & -0.00 & F0 & BF   & 99.99 & 20   & 79 & 1:13  & \\
 08092668-4914371 & 0.58 & 10.70 & 0.04 & 1.27 & 99.99 & BF   & 51 & 11   & 120 & 1:8  & \\
 08093053-4920443 & 0.81 & 10.90 & 0.01 & 0.07 & F0 & BF   & 52 & 13   & 104 & 1:14  & Broad lines \\
 08091401-4904029 & 0.84 & 11.23 & 0.14 & 0.16 & 99.99 & G    & 65 & 1   & 91 & 1:17  & Broad lines; ID 1 \\
 08104662-4917312 & 0.89 & 10.57 & 0.01 & 0.11 & 99.99 & G    & 34 & -5    & 31 & 3:5  & Low S/N \\
 08094893-4909370 & 0.93 & 10.74 & 99.99 & 99.99 & 99.99 & BF  & 4 & 10   & 79 & 1:11 & Broad lines \\
 08093815-4918403 & 0.96 & 10.62 & 0.11 & 0.95 & A7-A9 & BF   & 42 & 16   & 110 & 1:20a  & ID 2a \\ 
 08093815-4918403 & 0.96 & 10.62 & 0.11 & 0.95 & 99.99 & K    & 42 & 39   & 72 & 1:20b  & ID 2b \\
 08090287-4906022 & 1.02 & 11.60 & 0.02 & -0.07 & 99.99 & G    & 51 & 0.6  & 56 & 1:26  & \\
 08100087-4908324 & 1.04 & 10.80 & -0.01 & 0.01 & 99.99 & BF   & 44 & 13   & 31 & 1:9  & Emission? Low S/N \\
 08101352-4920438 & 1.10 & 10.78 & 0.00 & 0.03 & 99.99 & BF   & 22 & 26   & 15 & 1:10  & Low S/N \\
 08092293-4907575 & 1.14 & 12.00 & -0.06 & 0.60 & 99.99 & K    & 26 & 9   & 81 & 1:15  & \\
 08102774-4912095 & 1.15 & 10.81 & 0.04 & -0.14 & 99.99 & G    & 19 & -10.6 & 46 & 3:6  & \\
 08104984-4911258 & 1.16 & 11.58 & 0.06 & 0.14 & 99.99 & G    & 29 & 6   & 29 & 3:8  & \\
 08112074-4913100 & 1.18 & 11.85 & 0.02 & 99.99 & 99.99 & G    & 46 & 19   & 20 & 3:10  & \\
 08103470-4908399 & 1.30 & 12.20 & -0.03 & 0.05 & 99.99 & G    & 13 & 5   & 27 & 3:9  & \\
 08095109-4859022 & 1.37 & 12.96 & 0.10 & 99.99 & 99.99 & G    & 0 & 49   & 99 & 5:23  & \\
 08085790-4854189 & 1.38 & 99.99 & 0.02 & 99.99 & 99.99 & G    & 99.99 & 14   & 97 & 5:28  & \\
 08100380-4901071 & 1.49 & 12.36 & 0.02 & 0.02 & 99.99 & G    & 42 & 8   & 120 & 5:14  & Broad lines \\
 08093456-4920553 & 1.58 & 13.01 & 0.03 & 0.87 & 99.99 & K    & 40 & 27   & 190 & 2:13  & \\
 08090250-4858172 & 1.61 & 12.95 & 1.54 & 3.66 & 99.99 & G    & 99.99 & 14   & 62 & 5:11  & ID 8 \\
 08095421-4919220 & 1.62 & 12.50 & 0.01 & -0.09 & 99.99 & K    & 99.99 & 16   & 29 & 1:27  & \\
 08094117-4855146 & 1.63 & 99.99 & 0.03 & 99.99 & 99.99 & G    & 99.99 & 22   & 84 & 5:32  & \\
 08110234-4916249 & 1.65 & 12.90 & 0.01 & 0.79 & 99.99 & K    & 64 & 16   & 35 & 4:8  & \\
 08101542-4911095 & 1.81 & 13.03 & 0.18 & 99.99 & 99.99 & K    & 49 & 70   & 29 & 4:6  & \\
 08103793-4914225 & 1.87 & 13.20 & 0.11 & 99.99 & 99.99 & K    & 57 & 17   & 36 & 4:9  & \\
 08092276-4916309 & 1.91 & 13.42 & 0.07 & 99.99 & 99.99 & K    & 7 & 7   & 170 & 2:14  & Emission\\
 08101292-4914094 & 1.92 & 13.45 & 0.02 & 99.99 & 99.99 & K    & 99.99 & 13   & 190 & 2:9  & Emission \\
 08095274-4922170 & 1.97 & 13.33 & -0.06 & 99.99 & 99.99 & K    & 0 & 26   & 160 & 2:15  & \\
 08101799-4923502 & 2.09 & 13.71 & 0.05 & 99.99 & 99.99 & K    & 41 & 14   & 22 & 4:11  & Low S/N\\
 08091928-4904260 & 2.19 & 14.26 & 0.15 & 99.99 & K3e & K    & 99.99 & 21   & 26 & 5:36  & \\
 08110637-4912267 & 2.19 & 13.97 & 0.08 & 99.99 & 99.99 & K    & 36 & 23   & 22 & 4:13  & \\
 08092916-4901240 & 2.20 & 14.17 & 0.01 & 99.99 & 99.99 & K    & 6 & 3   & 29 & 5:10  & \\
 08091332-4857599 & 2.20 & 13.72 & 0.14 & 99.99 & 99.99 & K    & 99.99 & 0    & 40 & 5:33  & Emission?\\
 08092129-4900412 & 2.20 & 14.13 & 0.05 & 99.99 & 99.99 & K    & 59 & 1   & 38 & 5:35  & \\
 08085572-4857167 & 2.41 & 14.49 & 0.14 & 99.99 & 99.99 & K    & 99.99 & 31   & 31 & 5:17  & \\
 08104410-4913153 & 2.41 & 13.33 & 0.06 & 99.99 & 99.99 & K    & 71 & 68   & 37 & 4:7  & \\
 08101944-4907444 & 2.42 & 14.02 & 0.09 & 99.99 & 99.99 & K    & 66 & -30  & 170 & 2:17  & \\
 08092387-4915041 & 2.50 & 13.94 & 0.01 & 99.99 & 99.99 & K    & 0 & 22   & 130 & 2:24  & \\
 08101474-4912320 & 2.51 & 13.39 & 0.12 & 0.74 & 99.99 & K    & 3 & 110  & 220  & 2:7  & \\
 08090022-4920332 & 2.56 & 14.41 & 0.03 & 99.99 & 99.99 & K    & 40 & 9   & 110  & 2:20  & Emission \\
 08100398-4913027 & 2.65 & 13.87 & 0.14 & 99.99 & 99.99 & K    & 99.99 & 4   & 140  & 2:10a  & \\
 08094482-4902195 & 2.67 & 14.15 & 0.06 & 99.99 & 99.99 & K    & 99.99 & 54   & 40   & 5:29  & \\
 08100961-4915540 & 2.72 & 13.90 & 0.22 & 0.61 & 99.99 & K    & 48 & 8   & 170  & 2:8  &  \\
 08101083-4858477 & 2.74 & 14.41 & 0.06 & 99.99 & 99.99 & K    & 38 & 73   & 52   & 5:8  & \\
 08101987-4856000 & 2.75 & 14.46 & 0.05 & 99.99 & 99.99 & K    & 99.99 & 16   & 66   & 5:24  & Emission \\
 08100558-4914581 & 2.77 & 14.46 & 0.21 & 99.99 & 99.99 & K    & 26 & 35   & 140  & 2:11  & \\
 08091164-4857019 & 2.77 & 14.87 & 0.12 & 99.99 & 99.99 & K    & 99.99 & 70   & 24   & 5:20  & \\
 08092019-4907045 & 2.79 & 14.41 & 0.11 & 99.99 & 99.99 & K    & 0 & 33   & 150  & 2:16  & \\
 08101679-4913171 & 2.84 & 14.76 & -0.01 & 99.99 & 99.99 & K    & 0 & 91  & 110  & 2:18  & \\
 08092683-4900187 & 2.94 & 14.73 & 0.10 & 99.99 & 99.99 & K    & 0 & 50   & 30   & 5:19  & \\
 08101691-4856291 & 3.10 & 14.20 & 3.10 & 6.99 & 99.99 & K    & 99.99 & 29   & 64   & 5:12  & Broad lines; ID 9 \\
 08100955-4856343 & 3.11 & 15.31 & 0.06 & 99.99 & 99.99 & K    & 99.99 & 68   & 47   & 5:21  & \\
 08100893-4915413 & 3.12 & 15.29 & 0.15 & 99.99 & K7e & K    & 0 & -13   & 15   & 4:17  & Low S/N; emiss.? \\
 08100442-4850544 & 3.21 & 99.99 & 0.06 & 99.99 & 99.99 & K    & 99.99 & 26   & 17   & 5:38  &  \\
 08103186-4921296 & 3.25 & 15.04 & 0.14 & 99.99 & K7e & K    & 62 & -14   & 17   & 4:12  & Low S/N; emiss.? \\
 08100759-4900035 & 3.27 & 14.44 & 0.18 & 0.47 & 99.99 & K    & 69 & 136  & 49   & 5:25  & \\
 08095507-4858140 & 3.57 & 15.96 & 0.16 & 99.99 & 99.99 & K    & 99.99 & 15   & 23   & 5:22  & Emission \\
 08102318-4912086 & 3.61 & 15.99 & 0.20 & 99.99 & 99.99 & M    & 99.99 & 17   & 90   & 2:26  & Emission \\
 08091657-4909309 & 3.62 & 15.92 & 0.26 & 99.99 & M0e & M    & 99.99 & 11   & 84   & 2:21  & Emission \\
 08100749-4910447 & 3.72 & 16.09 & 0.07 & 99.99 & M0e & M    & 99.99 & 16   & 92   & 2:23  & Emission \\
 08091025-4902250 & 3.74 & 15.85 & 0.38 & 0.27 & 99.99 & M    & 0 & 10   & 18   & 5:18  & Emission \\
 08093105-4853166 & 3.75 & 99.99 & 0.04 & 99.99 & 99.99 & K    & 99.99 & 89   & 28 & 5:27  & \\ 
 08094556-4917364 & 3.90 & 16.12 & 0.16 & 99.99 & M0e & M     & 99.99 & 9   & 68   & 2:19  & Low S/N \\
 08102553-4906335 & 3.98 & 16.56 & 0.18 & 99.99 & M1e & M    & 99.99 & 16   & 80   & 2:28  & Emission \\
 08093685-4914214 & 4.00 & 15.74 & 0.27 & 99.99 & M1e & M    & 99.99 & 17   & 100  & 2:12  & Emission \\
 08102150-4901339 & 4.21 & 17.03 & 0.21 & 99.99 & 99.99 & M    & 99.99 & 10   & 27   & 5:30  & Low S/N; emiss.? \\
 08092437-4906282 & 4.27 & 17.04 & 0.19 & 99.99 & M2e & M    & 99.99 & 8   & 68   & 2:31  & Emission \\
 08092131-4905540 & 4.36 & 16.64 & 0.25 & 99.99 & 99.99 & M    & 99.99 & 13   & 86   & 2:25  & Emission \\
 08100180-4915454 & 4.42 & 17.38 & 0.19 & 99.99 & M1e & M    & 99.99 & 12   & 61   & 2:29  & Emission \\
 08102210-4911230 & 4.43 & 17.47 & 0.42 & 99.99 & M2e & M    & 99.99 & 22   & 64   & 2:32  & Emission \\
 08102103-4910448 & 4.64 & 17.81 & 0.08 & 99.99 & M3e & M    & 99.99 & 17   & 58   & 2:33  & Emission \\
 08093892-4915051 & 4.70 & 17.34 & 0.19 & 99.99 & 99.99 & M    & 99.99 & 16   & 61   & 2:22  & Emission \\
 08095429-4908418 & 4.71 & 18.27 & 0.21 & 99.99 & 99.99 & M    & 99.99 & 15   & 48   & 2:37  & Emission \\
 08085758-4911261 & 4.77 & 18.16 & 0.15 & 99.99 & 99.99 & M    & 99.99 & 12   & 42   & 2:34  & Emission \\
 08094273-4921168 & 4.85 & 18.64 & 0.22 & 99.99 & 99.99 & M    & 99.99 & 11   & 34   & 2:38  & \\
 08091111-4914183 & 5.03 & 18.19 & 0.39 & 99.99 & 99.99 & M    & 99.99 & 6   & 361 & 2:30  & \\
 08094499-4906241 & 5.06 & 18.04 & 0.72 & 99.99 & M4e & M    & 99.99 & 15   & 48   & 2:35  & \\
 08092592-4909585 & 5.07 & 18.45 & 0.03 & 99.99 & 99.99 & M    & 99.99 & 8   & 43   & 2:36  & Emission ? \\
 08100063-4853392 & 5.24 & 99.99 & 0.12 & 99.99 & 99.99 & M    & 99.99 & 8   & 16   & 5:31  & \\
 08094946-4916189 & 5.49 & 18.59 & 0.36 & 99.99 & 99.99 & M    & 99.99 & 17   & 47   & 2:27  & \\
 08093547-4913033 & 5.50 & 19.55 & 0.95 & 3.70 & M4.5 & M    & 99.99 & 11   & 28   & 2:6  & ID 7 \\
\enddata 
\tablenotetext{1}{Range of spectral types found in the \textit{VizieR}}
\tablenotetext{2}{Rough estimate of the spectral type for use with the corresponding radial velocity template,
a combination of the actual SpT and the airmass of observation (see \S \ref{sp})}
\tablenotetext{3}{Membership probability based on proper motion from \citet{Dias06}}
\tablenotetext{4}{Heliocentric velocity (as explained in \S \ref{sp})} 
\tablenotetext{5}{Signal-to-noise of spectrum estimated as the square root of counts in the continuum near 8490\AA\, }  
\tablenotetext{6}{Designation of spectrum based on the mask number (Table \ref{tablemasks}) and slit number within the mask}  
\tablenotetext{7}{ID for stars with 8 $\mu$m excess, their spectra shown on Fig. \ref{fig4}. Emission means emission inside Ca II NIR absorption lines} 
\tablecomments{Magnitudes de-reddened assuming uniform reddening of $A_{V}=0.19$. Where $V$ is unavailable,
we estimated $V-K$  from $I-K$ vs $V-K$ sequence constructed for cluster members. Objects sorted according to $(V-K)_{0}$. Radial
velocities and spectra are not reliable when S/N is low or lines are broad.} 
\end{deluxetable} 

\clearpage
\begin{deluxetable}{ccrrrrl} 
\tablecolumns{7}  
\tablewidth{0pt} 
\tabletypesize{\tiny} 
\tablecaption{NGC 2547 Members with mid-IR Excess\label{tableexc}}
\tablehead{ \colhead{2MASS Name} & \colhead{Sp. No} & \colhead{$(V-K)_{0}$} & \colhead{$(R-K)_{0}$} & 
\colhead{$(K-[8])_{0}$\tablenotemark{1}} & \colhead{$(K-[24])_{0}$\tablenotemark{1}} & \colhead{Remarks\tablenotemark{2}}
} 
\startdata
\sidehead{Selected based on 8 $\mu$m excess:} \\
  08091401-4904029 & 1:17         & 0.84 & 99.99 & \textbf{0.14} &  0.16          & ID 1 \\ 
  08093815-4918403 & 1:20a, 1:20b & 0.96 & 99.99 & \textbf{0.11} &  \textbf{0.95} & ID 2; EXC $K,[3]$\\ 
  08090720-4919144 & 99.99        & 2.76 &  2.04 & \textbf{0.26} & 99.99          & ID 4; EXC $[4],[5]$\\ 
  08093547-4913033 & 2:6          & 5.50 &  4.29 & \textbf{0.95} &\textbf{3.70}   & ID 7; EXC $[5]$\\  
  08090250-4858172 & 5:11         & 1.61 &  1.25 & \textbf{1.55} &\textbf{3.66}   & ID 8; EXC $[3],[4],[5]$\\
  08101691-4856291 & 5:12         & 3.10 &  2.47 & \textbf{3.10} & \textbf{6.99}  & ID 9; EXC $K,[3],[4],[5]$\\ 
\\
\sidehead{Selected based on 24 $\mu$m excess:} \\
   08092602-4911553 & 99.99        &-0.18 & 99.99 & 0.04          & \textbf{0.22} &            \\  
   08084981-4913437 & 99.99        &-0.04 & 99.99 & 0.02          & \textbf{0.17} &              \\   
   08100607-4914180 & 1:18         & 0.04 & 99.99 & 0.02          & \textbf{1.47} &              \\ 
   08112585-4912288 & 99.99        & 0.08 & 99.99 & -0.01         & \textbf{0.70} &               \\  
   08104233-4857253 & 99.99        & 0.12 & 99.99 & 0.00          & \textbf{0.28} &              \\   
   08100841-4900434 & 99.99        & 0.12 & 99.99 & 0.00          & \textbf{1.10} &               \\  
\\
   08110323-4900374 & 99.99        & 0.57 & 99.99 & 0.03          & \textbf{1.08} &            \\   
   08092668-4914371 & 1:8          & 0.58 & 99.99 & 0.04          & \textbf{1.27} &               \\ 
   08084571-4923473 & 99.99        & 0.62 & 99.99 & 0.00          & \textbf{0.96} &              \\   
   08111134-4904442 & 99.99        & 0.71 & 99.99 & 0.07          & \textbf{0.54} &              \\   
   08101673-4915173 & 99.99        & 0.72 & 99.99 & 0.03          & \textbf{0.79} &               \\  
\\
   08092293-4907575 & 1:15         & 1.14 & 99.99 & -0.06         & \textbf{0.60} &             \\ 
   08104546-4901068 & 99.99        & 1.23 & 99.99 & 0.05          & \textbf{0.43} &               \\  
   08101836-4906461 & 99.99        & 1.27 & 99.99 & 0.01          & \textbf{0.42} &               \\  
   08094507-4856307 & 99.99        & 1.27 & 99.99 & -0.01         & \textbf{0.45} &               \\  
   08085576-4923085 & 99.99        & 1.33 & 99.99 & 0.00          & \textbf{0.36} &               \\  
\\
   08090344-4859210 & 99.99        & 5.37 & 99.99 & 0.60          & \textbf{4.62} &       \\ 
\\
\sidehead{Selected based on possible 24 $\mu$m excess:} \\
   08104815-4923385 & 99.99        & 1.51 &  1.15 & -0.01         & \textbf{0.62} &               \\ 
   08093456-4920553 & 2:13         & 1.59 &  1.20 & 0.03          & \textbf{0.87} &               \\  
   08110009-4906442 & 99.99        & 1.59 &  1.23 & 0.03          & \textbf{0.84} &              \\  
   08110234-4916249 & 4:8          & 1.65 &  1.24 & 0.01          & \textbf{0.79} &              \\  
\\
   08101474-4912320 & 2:7          & 2.51 & 99.99 & 0.12          & \textbf{0.74} & RV \& PM NM \\ 
   08100961-4915540 & 2:8          & 2.72  & 2.12 & 0.22          & \textbf{0.61} &               \\ 
   08091758-4859252 & 99.99        & 2.67 &  2.11 & 0.09          & \textbf{0.69} &               \\  
\\
   08125256-4912123 & 99.99        & 99.99 & 2.21 & 99.99         & \textbf{1.48} &        \\ 
   08102491-4851482 & 99.99        & 99.99 & 1.83 & 0.12          & \textbf{1.90} & PM NM        \\  
\\
   08091770-4908344 & 99.99        & 5.56 & 4.30  & 0.64          & \textbf{3.18} &     \\    
\enddata 
\tablenotetext{1}{In bold are colors indicating excess} 
\tablenotetext{2}{RV: radial velocity; PM: proper motion; NM: non-member; EXC: ``excess also at'',
square brackets denote IRAC wavebands in $\mu$m}
\end{deluxetable}  

\begin{thebibliography}{} 

\bibitem[Allen \& Strom(1995)]{Allen95} Allen, L. E. \& Strom, K. M.  1995, \aj, 109, 1379
\bibitem[Baraffe et al.(2002)]{Baraffe02} Baraffe, I., Chabrier, G., Allard, F., \& Hauschildt, P. H.  2002, \aap, 382, 563 
\bibitem[Beichman et al.(2005)]{beichman05} Beichman, C.~A., et al.  2005, \apj, 626, 1061
\bibitem[Beichman et al.(2006)]{Beichman06} Beichman, C. A., Tanner, A., Bryden, G., Stapelfeldt, K. R.,
Werner, M. W., Rieke, G. H., Trilling, D. E., Lawler, S., \& Gautier, T. N.  2006, \apj, 639, 1166
\bibitem[Bessell \& Brett(1988)]{Bessell88} Bessell, M. S. \& Brett, J. M.  1988, \pasp, 100, 1134
\bibitem[Bonatto et al.(2006)]{Bonatto06} Bonatto, C., Bica, E., Ortolani, S., \& Barbuy, B.  2006, \aap, 453, 121
\bibitem[Bryden et al.(2006)]{Bryden06} Bryden, G. et al.  2006, \apj, 636, 1098
\bibitem[Buchanan et al.(2006)]{Buchanan06} Buchanan, C. L., Kastner, J. H., Forrest, W. J.,
Hrivnak, B. J., Sahai, R., Egan, M., Frank, A., \& Barnbaum, C.  2006, \aj, in press (astro-ph/0606756)
\bibitem[Burkert \& Ida(2006)]{Burkert06} Burkert, A. \& Ida, S.  2006, \apj, submitted (astro-ph/0608347)
\bibitem[Canup(2004)]{canup04} Canup, R.~M.  2004, \araa, 42, 441 
\bibitem[Carquillat et al.(1997)]{Carquillat97} Carquillat, J. M., Jaschek, C., Jaschek, M., \& Ginestet, N.  1997, \aaps, 123, 5
\bibitem[Chambers(2001)]{Chambers01} Chambers, J. E. 2001, Icarus, 152, 205 
\bibitem[Chen et al.(2006)]{Chen06} Chen, C. H. et al.  2006, \apjs, in press (astro-ph/0605277)
\bibitem[Clari\'{a}(1982)]{Claria82} Clari\'{a}, J. J.  1982, \aaps, 47, 323 
\bibitem[Currie et al.(2007a)]{Currie07} Currie, T., Kenyon, S., Rieke, G., Balog, Z., \& Bromley, B. C.  
  2007a, \apjl, in press (astro-ph/0706.0535)
\bibitem[Currie et al.(2007b)]{Curry06} Currie, T., Balog, Z., Kenyon, S. J., Rieke, G.,
 Prato, L., Muzerolle, J., \& Young, E. T.  2007b, \apj, 659, 599
\bibitem[D'Alessio et al.(1999)]{Dalessio99} D'Alessio, P., Calvet, N., Hartmann, L., Lizano, S., \& Cant\'{o}, J.
  1999, \apj, 527, 893
\bibitem[D'Antona \& Mazzitelli(1997)]{Dantona97} D'Antona, F. \& Mazzitelli, I.  1997,  Mem. Soc. Astron. Ital., 68, 807
\bibitem[de La Reza, Drake, \& da Silva(1996)]{deLaReza96} de La Reza, R., Drake, N. A., \& da Silva, L.  1996, \apj, 456, L115
\bibitem[Dias et al.(2006)]{Dias06}  Dias W.S., Florio, V., Assafin M., Alessi B.S., Libero V.  2006, \aap, 446, 949 
\bibitem[Dominik \& Decin(2003)]{Dominik03} Dominik, C. \& Decin, G.  2003, \apj, 598, 626
\bibitem[Drake et al.(2005)]{Drake05} Drake, N. A., de la Reza, R., da Silva, L., \& Lambert, D. L.  2002, \aj, 123, 2703 
\bibitem[Furlan et al.(2007)]{Furlan07} Furlan, E. et al.  2007, \apj in press (astro-ph/0705.0380)
\bibitem[Gorlova et al.(2004)]{Gorlova04} Gorlova, N. et al.  2004, \apjs, 154, 448 
\bibitem[Gorlova et al.(2006)]{Gorlova06} Gorlova, N., Rieke, G., Muzerolle, J., Stauffer, J., Siegler, N., 
 Young, E., \& Stansberry, J.  2006, \apj, in press (astro-ph/0606039)
\bibitem[Gutermuth et al.(2004)]{Gutermuth04} Gutermuth, R. A., Megeath, S. T., Muzerolle, J.,
 Allen, L. E., Pipher, J. L., Myers, P. C., \& Fazio, G. G.  2004, \apjs, 154, 374
\bibitem[Hanson et al.(1996)]{Hanson96} Hanson, M. M., Conti, P. S., \& Rieke, M. J.  1996, \apjs, 107, 281
\bibitem[Hartmann et al.(2005a)]{Hartmann05a} Hartmann, L., Megeath, S. T., Allen, L., Luhman, K., Calvet, N.,
D\'Alessio, P., Franco-Hernandez, R., \& Fazio, G.  2005a, \apj, 629, 881 
\bibitem[Hartmann et al.(2005b)]{Hartmann05b} Hartmann, L. et al. 2005b, \apj, 628, 147
\bibitem[Hern\'{a}ndez et al.(2005)]{Hernandez05} Hern\'{a}ndez, J. et al. 2005, \apj, 129, 856
\bibitem[Hern\'{a}ndez et al.(2006)]{Hernandez06} Hern\'{a}ndez, J., Brice\~{n}o, C., 
  Calvet, N., Hartmann, L., Muzerolle, J., \& Quintero, A.  2006, \apj, in press (astro-ph/0607562)
\bibitem[Hern\'{a}ndez et al.(2007)]{Hernandez07} Hern\'{a}ndez, J. et al.  2007, \apj, 662, 1067
\bibitem[Hillenbrand(2005)]{Hillenbrand05} Hillenbrand, L. A.  2005, "A Decade of Discovery: Planets Around Other Stars" 
  STScI Symposium Series 19, ed. M. Livio (astro-ph/0511083)
\bibitem[Ivanov et al.(2004)]{Ivanov04} Ivanov, V. D., Rieke, M. J., Engelbracht, C. W.,
 Alonso-Herrero, A., Rieke, G. H., \& Luhman, K. L.  2004, \apjs, 151, 387
\bibitem[Jeffries \& Tolley(1998)]{Jeffries98} Jeffries, R. D. \& Tolley, A. J.  1998, \mnras, 300, 331
\bibitem[Jeffries, Totten, \& James(2000)]{Jeffries00}  Jeffries, R. D., Totten, E. J., \& James, D. J.  2000, \mnras, 316, 950 
\bibitem[Jeffries et al.(2003)]{Jeffries03} Jeffries, R. D., Oliveira, J. M., Barrado y Navascu\'{e}s, D., \& Stauffer, J. R.
  2003, \mnras, 343, 1271
\bibitem[Jeffries et al.(2004)]{Jeffries04} Jeffries, R. D., Naylor, T., Devey, C. R., Totten, E. J.  2004, \mnras, 351, 1401
\bibitem[Jeffries \& Oliveira(2005)]{Jeffries05} Jeffries, R. D. \& Oliveira,  J. M.,  2005, \mnras, 358, 13
\bibitem[Jeffries et al.(2006)]{Jeffries06}  Jeffries, R. D.,  Evans, P. A., Pye, J. P., \& Briggs, K. R.  2006,
\mnras, 367, 781
\bibitem[Jura(1990)]{Jura90} Jura, M.  1990, \apj, 365, 317
\bibitem[Kenyon \& Hartmann(1995)]{Kenyon95} Kenyon, S. J. \& Hartmann, L.  1995, \apjs, 101, 117
\bibitem[Kenyon \& Bromley(2004a)]{Kenyon04a} Kenyon, S. J. \& Bromley, B. C.  2004a, \apj, 602, L133
\bibitem[Kenyon \& Bromley(2004b)]{Kenyon04b} Kenyon, S. J. \& Bromley, B. C.  2004b, \aj, 127, 513
\bibitem[Kenyon \& Bromley(2005)]{Kenyon05} Kenyon, S. J. \& Bromley, B. C.  2005, \aj, 130, 269
\bibitem[Kenyon \& Bromley(2006)]{Kenyon06} Kenyon, S. J. \& Bromley, B. C.  2006, \aj, 131, 1837
\bibitem[Kim et al.(2005)]{Kim05} Kim, J. S., et al.  2005, \apj, 632, 659 
\bibitem[Kleine et al.(2002)]{Kleine02} Kleine, T., Munker, C., Mezger, K., \& Palme, H.  2002, \nat, 418, 952
\bibitem[Kleinmann et al.(1986)]{KH86} Kleinmann, S. G. \& Hall, D. N. B.  1986, \apjs, 62, 501
\bibitem[Kleinmann et al.(1986)]{Kleinmann86} Kleinmann S. G., Cutri R. M., Young E. T., Low F. J., \& Gillett F.C.  1986, 
IRAS Serendipitous Survey Catalog (On-line VizieR catalog II/126)
\bibitem[Lada et al.(2006)]{Lada06} Lada, C. J. et al.  2006, \aj, 131, 1574
\bibitem[Littlefair et al.(2003)]{Littlefair03} Littlefair, S. L., Naylor, T., Jeffries, R. D., Devey, C. R., \& Vine, S.
 2003, \mnras, 345, 1205
\bibitem[Low et al.(2005)]{Low05} Low, F. J., Smith, P. S., Werner, M., Chen, C., Krause, V., Jura, M., 
\& Hines, D. C.  2005, ApJ, 631, 1170
\bibitem[Lyra et al.(2006)]{Lyra06} Lyra, W., Moitinho, A., van der Bliek, N. S., \& Alves, J.  2006, \aap, 453, 101
\bibitem[Mamajek et al.(2004)]{Mamajek04} Mamajek, E. E., Meyer, M. R., Hinz, P. M., Hoffmann, W. F., Cohen, M., Hora, J. L.
  2004, \apj, 612, 496
\bibitem[Mathis (1990)]{Mathis90} Mathis, J. S.  1990, \araa, 28, 37
\bibitem[Megeath et al.(2005)]{Megeath05} Megeath, S. T., Hartmann, L., Luhman, K. L., \& Fazio, G. G.  2005, \apj, L634, 113
\bibitem[Meixner et al.(2006)]{Meixner06} Meixner, M. et al.  2006, \apj, in press (astro-ph/0606356)  
\bibitem[Meyer et al.(2006)]{Meyer06} Meyer, M. R, Backman, D. E., Weinberger, A. J., \& Wyatt, M. C.  2006, 
 in ``Protostars and Planets V'', Univ. of Arizona Press, editors: B. Reipurth, D. Jewitt, and K. Keil 
\bibitem[Meyer et al.(2007)]{Meyer07} Meyer, M. R. et al.  2007, PASP, in press (astro-ph/0701058)
\bibitem[Moorwood et al.(1998)]{Moorwood98} Moorwood, A., et al. 1998, Messenger, 94, 7
\bibitem[Moro-Mart\'{i}n \& Malhotra(2002)]{MoroMartin02} Moro-Mart\'{i}n, A. \& Malhotra, R.  2002, \aj, 124, 2305
\bibitem[Munari \& Tomasella(1999)]{Munari99} Munari, U. \& Tomasella, L.  1999, \aaps, 137, 521 
\bibitem[Naylor et al.(2002)]{Naylor02} Naylor, T., Totten, E. J., Jeffries, R. D., Pozzo, M., Devey, C. R., \& Thompson, S. A.,
  2002, \mnras, 335, 291
\bibitem[Naylor \& Jeffries(2006)]{Naylor06} Naylor, T. \& Jeffries, R. D. 2006, \mnras, 373, 1251
\bibitem[Oliveira et al.(2003)]{Oliveira03} Oliveira, J. M., Jeffries, R. D., Devey, C. R.,
  Barrado y Navascu\'{e}s, D., Naylor, T., Stauffer, J. R., \& Totten, E. J.  2003, \mnras, 342, 651
\bibitem[Oudmaijer et al.(1995)]{Oudmaijer95} Oudmaijer, R. D., Waters, L. B. F. M., van der Veen, W. E. C. J.,
  \& Geballe, T. R.  1995, \aap, 299, 69
\bibitem[Padgett et al.(2006)]{Padgett06} Padgett, D. L. et al.  2006, \apj, 645, 1283
\bibitem[Pascucci et al.(2006)]{Pascucci06} Pascucci, I. et al.  2006, \apj, 651, 1177 
\bibitem[Patten et al.(2006)]{Patten06} Patten, B. M., et al.  2006, \apj, 651, 502
\bibitem[Plets et al.(1997)]{Plets97} Plets, H., Waelkens, C., Oudmaijer, R. D., \& Waters, L. B. F. M.  1997, \aap, 323, 513 
\bibitem[Rhee et al.(2007)]{Rhee07} Rhee, J. H., Song, I., \& Zuckerman, B.  2007, \apj, in press (astro-ph/0706.1265)
\bibitem[Riaz, Gizis, \& Hmiel(2006)]{Riaz06} Riaz, B., Gizis, J. E., \& Hmiel, A.  2006, \apj, 639, L79
\bibitem[Rieke et al.(2005)]{Rieke05} Rieke, G. H., et al.  2005, \apj, 620, 1010
\bibitem[Robichon et al.(1999)]{Robichon99} Robichon., N., Arenou, F., Mermilliod, J.-C., Turon, C.  1999,  \aap, 345, 471
\bibitem[Schiavon et al.(1997)]{Schiavon97} Schiavon, R. P., Barbuy, B., Rossi, S. C. F., \& Milone, A.
\bibitem[Sicilia-Aguilar et al.(2006)]{Sicilia06} Sicilia-Aguilar, A. et al.  2006, \apj, 638, 897
\bibitem[Sicilia-Aguilar et al.(2007)]{Sicilia07} Sicilia-Aguilar, A. et al.  2007, \apj, 659, 1637
\bibitem[Siegler et al.(2006)]{Siegler06} Siegler, N., Muzerolle, J., Mamajek, E., Young, E., Rieke, G., 
 Trilling, D., Gorlova, N., \& Su, K. Y. L.  2006, \apj, in press
\bibitem[Siess et al.(2000)]{Siess00}  Siess L., Dufour E., Forestini M. 2000, \aap, 358, 593
\bibitem[Silverstone et al.(2006)]{Silverstone06} Silverstone, M. D., et al.  2006, \apj, 639, 1138
\bibitem[Su et al.(2006)]{Su06} Su, K. Y. L. et al.  2006, \apj, submitted
\bibitem[Torres-Dodgen \& Weaver(1993)]{Torres93} Torres-Dodgen, A. V. \& Weaver, W. B.  1993, \pasp, 105, 693
\bibitem[Uzpen et al.(2005)]{Uzpen05} Uzpen, B., et al. 2005, \apj, 629, 512
\bibitem[Uzpen et al.(2006)]{Uzpen06} Uzpen, B., et al. 2006, \apj, in press (astro-ph/0612235)
\bibitem[Wallace \& Hinkle(1997)]{Wallace97} Wallace, L. \& Hinkle, K. 1997, \apjs, 111, 445
\bibitem[White \& Hillenbrand(2005)]{White05} White, R. J \& Hillenbrand, L. A.  2005, \apj, 621, L65
\bibitem[Worley \& Douglass(1997)]{Worley96} Worley C., E. \& Douglass G., G. 1997, \aaps, 125, 523 (On-line VizieR catalog  I/237)
\bibitem[Young et al.(2004)]{Young04} Young, E. T., et al.  2004, \apjs, 154, 428
\bibitem[Zacharias et al.(2003)]{ucac2} Zacharias N., Urban S. E., Zacharias M. I., Wycoff G. L., Hall D. M., 
     Germain M. E., Holdenried E.R., \& Winter L.  2003,  The Second U.S. Naval Observatory CCD 
     Astrograph Catalog
\bibitem[Zuckerman, Kim \& Liu(1995)]{Zuckerman95} Zuckerman, B., Kim, S. S., \& Liu, T.  1995, \apj, 446, L79

\end{thebibliography}
\end{document}